\documentclass[twocolumn,preprintnumbers]{revtex4}
\usepackage{graphicx,subfigure}%
\usepackage{amsmath}

\usepackage{amsmath}
\usepackage{dcolumn}
\usepackage{bm}
\usepackage{epsfig}
\usepackage{epstopdf}
\usepackage{graphicx}
\usepackage{amsfonts}
\usepackage{amssymb}
\usepackage{mathrsfs}
\usepackage{color}
\usepackage{multirow}
\usepackage{appendix}

\begin{document}

\title{Single-qubit measurement of Heisenberg uncertainty lower bounds for three incompatible observables}
\author{K. Rehan$^{1,3}$}
\author{T. P. Xiong$^{4}$}
\author{L.-L. Yan$^{2}$}
\email{llyan@zzu.edu.cn}
\author{F. Zhou$^{1}$}
\email{zhoufei@wipm.ac.cn}
\author{J. W. Zhang$^{1,3}$}
\author{J. C. Li$^{1,3}$}
\author{L. Chen$^{1}$}
\author{W. L. Yang$^{1}$}
\author{M. Feng$^{1,5}$}
\email{mangfeng@wipm.ac.cn}
\affiliation{$^{1}$ State Key Laboratory of Magnetic Resonance and Atomic and Molecular Physics,
Wuhan Institute of Physics and Mathematics, Chinese Academy of Sciences, Wuhan, 430071, China\\
$^{2}$ School of Physics and Engineering, Zhengzhou University, Zhengzhou 450001, China \\
$^{3}$ School of Physics, University of the Chinese Academy of Sciences, Beijing 100049, China\\
$^{4}$ Guangxi Key Laboratory of Trusted Software, Guilin University of Electronic Technology, Guilin, 541004,China\\
$^{5}$ Department of Physics, Zhejiang Normal University, Jinhua 321004, China}

\begin{abstract}
Being one of the centroidal concepts in quantum theory, the  fundamental constraint imposed by Heisenberg uncertainty relations has always been a subject of immense attention and challenging in the context of joint measurements of general quantum mechanical observables. In particular, the recent extension of the original uncertainty relations has  grabbed a distinct research focus and set a new ascendent target in quantum mechanics and quantum information processing. In the present work we explore the joint measurements of three incompatible observables, following the basic idea of a newly proposed error trade-off relation.
In comparison to the counterpart of two incompatible observables, the joint measurements of three incompatible observables are more complex and of more primal interest in understanding quantum mechanical measurements. Attributed to the pristine idea proposed by Heisenberg in 1927, we develop the error trade-off relations for compatible observables to categorically approximate the three incompatible observables. Implementing these relations we demonstrate the first experimental witness of the joint measurements for three incompatible observables using a single ultracold $^{40}Ca^{+}$ ion in a harmonic potential. We anticipate that our inquisition would be of vital importance for quantum precision measurement and other allied quantum information technologies.

\end{abstract}

\maketitle

\section{introduction}
Uncertainty principle, as one of the fundamental postulations in quantum theory, was first introduced by Heisenberg, which raised error bounds for joint measurements of non-commuting observables in preparing/measuring quantum states \cite{heisenberg}.
But the original forms of uncertainty relations were derived by Kennard \cite{kennard}, Weyl \cite{weyl} and Robertson \cite{robertson}, which are mathematical descriptions of deviation in the measurement statistics of two incompatible observables. Physically, these forms of uncertainty relation concern separate measurements of the two incompatible observables performed on two ensembles of identically prepared quantum systems, which is conceptually different from the Heisenberg's idea of a trade-off for the errors of approximate simultaneous or
successive measurements performed on the same system \cite{peres}.

\begin{figure}[hbtp]
\centering {\includegraphics[width=8 cm, height=5.9 cm]{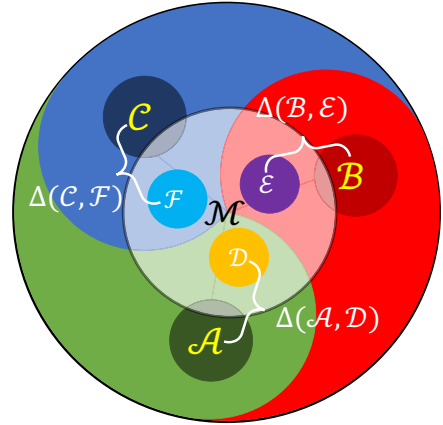}}
\caption{(Color online) Schematic for verifying lower bounds of the Heisenberg uncertainty relations for three incompatible observables by triplewise joint measurements. The quantum apparatus can measure the incompatible observables $\mathcal{A}$, $\mathcal{B}$ and $\mathcal{C}$ individually, while not simultaneously. Three compatible observables $\mathcal{D}$, $\mathcal{E}$, and $\mathcal{F}$ are employed to approximate
$\mathcal{A}$, $\mathcal{B}$ and $\mathcal{C}$, respectively, during the joint measurement by $\mathcal{M}$.}
\label{Fig1}
\end{figure}

From the operational aspect, measuring two incompatible observables $A$ and $B$ inevitably involves disturbance, e.g., a single measurement of $A$ can be accomplished with an imminent disturbance in any subsequent  or simultaneous measurement of $B$. As a result, treatment of the inaccuracy in an approximate measurement of $A$ and the disturbance of a subsequent or simultaneous measurement of $B$ necessitates a trade-off, which is known as an error-disturbance relation. Based on this idea, new inequalities for uncertainty relations have been independently proposed over past years \cite{tho1,tho2,tho3,tho4,tho5,tho6,tho7,prl-111-160405,PRA-89-012129}. Although some of those inequalities were verified experimentally \cite{exp1,exp2,exp3,exp4,exp5,exp6,exp7,exp8}, the debates still have been lasting due to disagreements on appropriate quantification of error and disturbance. In particular, some authors of the present work have recently worked on experimental verification of a trade-off proposed by Busch, Lahti and Werner (BLW) in \cite{PRA-89-012129},
which employed two compatible observables $C$ and $D$ to approximate, respectively, the incompatible observables $A$ and $B$, following the quintessence of Heisenberg's
original idea in 1927. Precise lower bounds of the uncertainty relations were witnessed, in conformity with the predictions in \cite{PRA-89-012129}, by simultaneously detecting incompatible observables encoded in a single ultracold trapped ion \cite{SA-2-e1600578,njp-19-063032}.

The present work concerns a more complicated situation, that is, triplewise joint measurements of three incompatible observables in a qubit. For a single qubit, e.g., a spin-1/2 particle, the spin operators  given by three Pauli operators $\sigma_{k}~(k=x,y,z)$, constitute the fundamental representation of SU(2), describing completely the properties of the particle. Consequently, compared with the counterpart regarding two incompatible observables, joint measurements of three incompatible observables encoded in a qubit should be of more fundamental interest. We have noticed the previous attempts to explore the uncertainty principle encompassing three or more observables \cite{multi-v1,multi-v2,multi-v3,multi-v4,multi-v5,multi-v6,multi-v7,multi-v8,multi-v9,multi-v10, multi-v11,multi-v101,multi-v12,multi-v13,multi-v14,multi-v104,multi-v15}, where basically different relations of the uncertainty principle, based on preparation and entropy, have been investigated. However, none of them is relevant to the error-disturbance relation. For example, a recent experiment \cite{multi-v15}, using a single spin inside the diamond nitrogen-vacancy center, demonstrated the preparation uncertainty for the triple components of the angular momentum.

In the present paper, we report the first experimental confirmation of lower bounds of the error-disturbance relations by extending the BLW proposal in \cite{PRA-89-012129} to three compatible observables approximating, respectively, three incompatible observables (See Fig. \ref{Fig1}). Since the prevailing situation is much more complicated than considered in \cite{PRA-89-012129}, we will only focus on some typical cases, e.g., three incompatible observables to be orthogonal or coplanar, which provide more alluring physical pictures than the general cases for the concerned problem without any resultant damage in the context of our insight into the problem's essence. Our experiment utilizes a single qubit encoded in a trapped
ultracold $^{40}Ca^{+}$ ion, which could be manipulated by high-level control of lasers. As will be presented below, by unitary operations under carrier transitions, we witness the lower bounds of the Heisenberg uncertainty relations from the triplewise joint measurements and discover some novel characteristics completely different from the usual considerations with two incompatible observables.

The paper is organized as follows. We first describe the theoretical scheme regarding the joint measurements of three incompatible observables, and then present experimental measurements in verifying the predicted lower bounds. A concise discussion and a short summary are given before the end of the main text. Some details of the theoretical derivations and experimental implementation can be found in Appendix.

\section{Theoretical scheme}

In this work we consider three incompatible observables $\mathcal{A}$, $\mathcal{B}$ and $\mathcal{C}$, and explore their uncertainty relations under triplewise joint measurements. This basically refers to extending the BLW proposal of pairwise joint measurements, as detailed in \cite{PRA-89-012129}, to the triplewise counterparts, as sketched in Fig. \ref{Fig1}, where three compatible observables $\mathcal{D}$, $\mathcal{E}$, $\mathcal{F}$ represent the corresponding approximations of the target observables $\mathcal{A}$, $\mathcal{B}$, $\mathcal{C}$.
For clarity, we first introduce the three compatible observables as well as their measurements. Then three incompatible observables, as the targets of the three compatible observables to approach are elucidated.

\subsection{Joint measurements of three compatible observables}

An arbitrary observable $\mathcal{O}$ for a qubit can be written as $\mathcal{O}=\bm{o}\cdot\bm{\sigma}$, where $\bm{o}=(o_x,o_y,o_z)$ is a general real vector with $|\bm{o}|\leq 1$ and $\bm{\sigma}=(\sigma_x,\sigma_y,\sigma_z)$ is the vector regarding Pauli operators. The observable $\mathcal{O}$ has two positive operators $O_{\pm}=(1\pm \bm{o}\cdot\bm{\sigma})/2$.

Three observables $\mathcal{D}$, $\mathcal{E}$ and $\mathcal{F}$  are called $triplewise \ jointly \ measurable$ if there is a joint measurement observable $\mathcal{M}$ with eight measurement operators $M_{\bm{\mu}}$ ($=M_{\mu_1 \mu_2 \mu_3}$ with $\mu_1,\mu_2,\mu_3=\pm$) making the three given observables as the marginals, e.g., $D_{\mu_1}=\sum_{\mu_2 \mu_3}M_{\mu_1 \mu_2 \mu_3}$. This implies that the observables $\mathcal{D}$, $\mathcal{E}$ and $\mathcal{F}$ can be indirectly measured by means of measuring $\mathcal{M}$, as detailed later.

Three observables to be triplewise jointly measurable should satisfy \cite{arXiv-1313-6470v1},
\begin{equation}
\sum_{k=0}^3|\bm{\Lambda}_k-\bm{\Lambda}_{\text{FT}}|\leq 4,
\label{Eq1}
\end{equation}
where $\bm{\Lambda}_{\text{FT}}$ denotes the Fermat-Toricelli (FT) vector of four vectors $\bm{\Lambda}_0=\sum_{k=1}^3\bm{\lambda}_k$ and $\bm{\Lambda}_k=2\bm{\lambda}_k-\bm{\Lambda}_0 \ (k=1,2,3)$ with $\bm{\lambda}_{1,2,3}$ the vectors regarding the observables $\mathcal{D}$, $\mathcal{E}$ and $\mathcal{F}$, respectively. The FT vector for a set of three or more vectors $\bm{v}_a$ in Euclidean space, which is unique and always exists, represents the vector $\bm{v}$ with the total distance $\sum_a|\bm{v}_a-\bm{v}|$ minimized \cite{EMAA-10-53}.
As such, the measurement operators of $\mathcal{M}$ for three triplewise joint measurement observables take the following form \cite{arXiv-1313-6470v1},
\begin{equation}
M_{\bm{\mu}}=\frac{1}{8}(I+\sum_{i>j}\mu_i\mu_jZ_{ij}+\sum_{i=1}^3\mu_i\bm{\lambda}_i\cdot\bm{\sigma}-\mu_1\mu_2\mu_3\bm{\Lambda}_{\text{FT}}\cdot\bm{\sigma}),
\label{Eq2}
\end{equation}
where $Z_{ij}=1-(|\bm{\Lambda}_i-\bm{\Lambda}_{\text{FT}}|-|\bm{\Lambda}_j-\bm{\Lambda}_{\text{FT}}|)/2$ with $i>j$ and $i,j=1,2,3$.

In order to understand Eq. (\ref{Eq2}) in a more simplified way, we take into account two special cases as explained below. The first case encapsulates the situation of three observables existing in mutual orthogonality, for which the satisfying condition in Eq. (\ref{Eq1}) reduces to \cite{arXiv-1313-6470v1,PRD-33-2253,PRA-76-052313},
\begin{equation}
\sum_i|\bm{\lambda}_i|^2\leq 1.
\label{Eq3}
\end{equation}
We have to mention that the joint measurement operators given by Eq. (\ref{Eq2}) are incompact although they satisfy Eq. (\ref{Eq3}). A proper compact form of $\mathcal{M}$ could be written as \cite{PRD-33-2253,PRA-76-052313},
\begin{equation}
M_{\bm{\mu}}=\frac{1}{8}(I+\sum_{i}\mu_i\bm{\lambda}_i\cdot\bm{\sigma}),
\label{Eq4}
\end{equation}
where $\bm{\mu}=(\mu_1,\mu_2,\mu_3)$ with $\mu_{i}=\pm$ ($i=1,2,3$).

The second special case deals with a reduction from three observables to two. Normally, a pair of observables $\mathcal{D}$ and $\mathcal{E}$ defined by two vectors $\bm{\lambda}_{1}$ and $\bm{\lambda}_{2}$ are $jointly \ measurable$ if and only if \cite{PRD-33-2253,njp-19-063032,SA-2-e1600578,PRA-89-012129}
\begin{equation}
|\bm{\lambda}_1+\bm{\lambda}_2|+|\bm{\lambda}_1-\bm{\lambda}_2|\leq 2,
\label{Eq5}
\end{equation}
whose joint measurement operators are
\begin{equation}
M_{\bm{\mu}}=\frac{1}{4}(G I+\sum_{i}\mu_i\bm{\lambda}_i\cdot\bm{\sigma}),
\label{Eq6}
\end{equation}
such that $\bm{\mu}=(\mu_1,\mu_2)$ and $G=1+\mu_1\mu_2\bm{\lambda}_1\cdot\bm{\lambda}_2$  with $\mu_{i}=\pm$ ($i=1,2$). This is actually the case investigated previously in \cite{PRA-89-012129,SA-2-e1600578}.

\subsection{Uncertainty relations for three incompatible observables}


In general, the three incompatible observables $\mathcal{A}$, $\mathcal{B}$ and $\mathcal{C}$ do not satisfy the triplewise jointly measurable condition under the limitation imposed by Eq. (\ref{Eq1}) unless they are colinear. In order to undertake a quantitative treatment, we consider $\mathcal{A}=\bm{a}\cdot\bm{\sigma}$, $\mathcal{B}=\bm{b}\cdot\bm{\sigma}$ and $\mathcal{C}=\bm{c}\cdot\bm{\sigma}$ with $|\bm{a}|=|\bm{b}|=|\bm{c}|=1$. Then
we assume the three compatible observables to be $\mathcal{D}=\bm{d}\cdot\bm{\sigma}$, $\mathcal{E}=\bm{e}\cdot\bm{\sigma}$ and $\mathcal{F}=\bm{f}\cdot\bm{\sigma}$ with $|\bm{d}|,|\bm{e}|, |\bm{f}|\leq 1$ as the approximations of $\mathcal{A}$, $\mathcal{B}$ and $\mathcal{C}$, respectively. By extending the trade-off idea in \cite{PRA-89-012129}, we define the state-independent uncertainty relation under the triplewise joint measurements as
\begin{eqnarray}
&&\Delta(\mathcal{A},\mathcal{B},\mathcal{C})\equiv\Delta(\mathcal{A},\mathcal{D})+\Delta(\mathcal{B},\mathcal{E})+\Delta(\mathcal{C},\mathcal{F}) \notag \\
&&\ \ =\max_{\rho}[\Delta_{\rho}(\mathcal{A},\mathcal{D})]+\max_{\rho}[\Delta_{\rho}(\mathcal{B},\mathcal{E})]+\max_{\rho}[\Delta_{\rho}(\mathcal{C},\mathcal{F})] \notag \\
&&\ \ =2(|\bm{a}-\bm{d}|+|\bm{b}-\bm{e}|+|\bm{c}-\bm{f}|),
\label{Eq7}
\end{eqnarray}
which ascribes the new Heisenberg uncertainty relation we obtain and would be consequently verified later on by the trapped-ion system. The Wasserstein distances (of order 2) between the measurement probabilities in Eq. (\ref{Eq7}), i.e.,
\begin{equation}
\Delta_{\rho}(\mathcal{X},\mathcal{Y})=2\sum_{\mu=\pm}|p_{\rho}^{X_{\mu}}-p_{\rho}^{Y_{\mu}}|=2|(\bm{x}-\bm{y})\cdot\bm{r}|,
\label{Eq8}
\end{equation}
define the state-dependent uncertainty relations \cite{PRA-89-012129}, in which $\mathcal{X}=\mathcal{A},\mathcal{B},\mathcal{C}$, $\mathcal{Y}=\mathcal{D},\mathcal{E},\mathcal{F}$ and the qubit state $\rho=(1+\bm{r}\cdot\bm{\sigma})/2$ with $ |\bm{r}|=1 $. $p_{\rho}^{K_{\mu}}$ denotes the measurement probability obtained by measuring $K_{\mu}$ on the state $\rho$, i.e., $p_{\rho}^{K_{\mu}}=\text{Tr}[K_{\mu}\rho]$ with $K=X,Y$.

For the set $\Xi$ consisting of all the groups of three observables satisfying the triplewise joint measurement condition, the lower bound of the uncertainty relation, based on the minimization of the vector distances regarding the incompatible and compatible observables, is defined as
\begin{equation}
\Delta_{lb}=\min_{(\bm{d},\bm{e},\bm{f})\in \Xi} 2(|\bm{a}-\bm{d}|+|\bm{b}-\bm{e}|+|\bm{c}-\bm{f}|).
\label{Eq9}
\end{equation}
Thus the uncertainty relation can be simply expressed as
\begin{equation}
\Delta(\mathcal{A},\mathcal{B},\mathcal{C})\geq \Delta_{lb}.
\label{Eq10}
\end{equation}
However, in contrast to the pairwise joint measurements with some analytical results for the optimal approximation, solving Eq. (\ref{Eq9}) analytically is difficult. As a result, we try below numerical solutions to Eq. (\ref{Eq9}) for most cases.
Our experimental observations, as presented later, are also based on Eq. (\ref{Eq9}) by employing a penalty function method (See Appendix A for details).

\section{experimental system and basic operations}


Our experiment is carried out on a single ultracold $^{40}Ca^{+}$ ion confined in a linear Paul trap as employed previously \cite{SA-2-e1600578,njp-19-063032}, which is constituted by four parallel blade-like electrodes and two end-cap electrodes. Constant voltage applied to the end-caps ensures axial confinement and the rf potential applied to the blade electrodes via helical resonator corresponds to $\Omega_{rf}/2\pi = 20.6$ MHz with the power of 5.5 W. Under the pseudo-potential approximation, we have the axial and radial frequencies of the trap potential to be, respectively, $\omega_z/2\pi=1.0$ MHz and $\omega_r/2\pi=1.2$ MHz. To contrive an intrinsic Zeeman substructure and hence to ascertain a peculiar quantization axis, we employ a magnetic field of 0.6 mT directed in axial orientation, yielding the ground state $4^2S_{1/2}$ and the metastable state $3^2D_{5/2}$ split into two and six hyperfine energy levels, respectively. We encode qubit  $\mid\downarrow\rangle$ in $|4^{2}S_{1/2}, m_{J}=+1/2\rangle$ and $\mid\uparrow\rangle$ in $|3^{2}D_{5/2}, m_{J}=+3/2\rangle$ with $m_{J}$ being the magnetic quantum number.

After Doppler cooling and resolved sideband cooling, the $z$-axis motional mode of the ion is cooled down to its vibrational ground state
with the final average phonon number of $0.030(7)$. A narrow-linewidth 729-nm laser radiates the ultracold ion with an incident angle of $22.5^{\circ} $ between the laser and the $z$-axis, yielding the carrier-transition Hamiltonian $H_c=\Omega(\sigma_+e^{i\phi_{L}}+\sigma_-e^{-i\phi_{L}})/2$, with the Rabi frequency $\Omega$ as the laser-ion coupling strength in units of $\hbar=1$ and the laser phase $\phi_{L}$. The system evolution under the execution of the carrier-transition operator is given by \cite{SA-2-e1600578},
\begin{equation}
U_{C}(\theta_{L},\phi_{L}) = \cos\frac{\theta_{L}}{2}I - i\sin\frac{\theta_{L}}{2}(\sigma_{x}\cos\phi_{L} -\sigma_{y}\sin\phi_{L}),
\label{Eq25}
\end{equation}
where the parameter $\theta_{L}=\Omega t$ is determined by the evolution time and $\Omega/2\pi=47.0(5)$ kHz throughout our experiment. Since the inherent decay and dephasing times of the qubit are, respectively, 1.1 s and 2 ms, much longer than the operation time, we can consider to work in an isolated quantum system.

Prior to proceeding to the substantial experimental operations to witness the uncertainty relation lower bounds, we are required to prepare an optimal state $\rho_{\text{op}}$ which is to maximize the state-dependent uncertainty $\Delta_{\rho}(\mathcal{A},\mathcal{D})$  with $\bm{r}=(\bm{a}-\bm{d})/|\bm{a}-\bm{d}|$. Starting from this state, we check Eq. (\ref{Eq7}) experimentally.
The required operations include execution of coupling of $|\uparrow\rangle$ and $|\downarrow\rangle$ by a 729-nm laser under a unitary evolution $U_{C}(\theta_{L1},\phi_{L1})$ (defined below) and measurement of the necessary observables $A$, $B$, $C$ and $M$ under another evolution $U_{C}(\theta_{L2},\phi_{L2})$ (defined below). Finally, detection is made on the state $|\uparrow\rangle$.

The preparation of an optimal state is achieved by the unitary operator in Eq. (\ref{Eq25}) with $\rho_{\text{op}}=U_{C}(\theta_{L1},\phi_{L1})\mid\downarrow\rangle\langle\downarrow\mid U^{\dagger}_{C}(\theta_{L1},\phi_{L1})$, that is,
\begin{equation}
r_x=\sin\theta_{L1}\sin\phi_{L1},\ r_y=\sin\theta_{L1}\cos\phi_{L1}, r_z=-\cos\theta_{L1}.
\label{Eq28}
\end{equation}
where the angles will be further clarified later.

The measurement operator $M$ is a positive operator-valued measure. As proven in \cite{SA-2-e1600578}, such a positive operator-valued measure with rank of $1$ can be measured by a single qubit through unitary transformations, at the expense of losing generality, i.e., no possibility to observe the region above the lower bound. Nevertheless, in
this way, the lower bound, corresponding to optimal approximations, can be observed as desired in a single qubit.
As such, the measurement operator $M=\text{Tr[M]}(1+\bm{m}\cdot\bm{\sigma})/2$ with $ \bm{m} $ having rank $1$ can be rewritten as $M=\text{Tr}[M] U_{C}^{\dagger}(\theta_{L2},\phi_{L2})\mid\uparrow\rangle\langle\uparrow\mid U_{C}(\theta_{L2},\phi_{L2})$ \cite{SA-2-e1600578}. Thus we have
\begin{equation}
m_x=\sin\theta_{L2}\sin\phi_{L2},\ m_y=\sin\theta_{L2}\cos\phi_{L2}, m_z=\cos\theta_{L2}.
\label{Eq26}
\end{equation}
For convenience of the experimental implementation in our system, we choose $\theta_{L2}=\arccos m_z$, which means $\sin(\theta_{L2})\geq 0$. Thus, $ \phi_{L2}$, which only depends on $m_x$ and $m_y$, is given by,
\begin{equation}
 \phi_{L2}=\frac{\pi}{2}(1-\text{sign}(m_y))+\arctan \frac{m_x}{m_y},
 \label{Eq27}
\end{equation}
where $\text{sign}(m_y)=1$ if $m_y>0$ and $\text{sign}(m_y)=-1$ if $m_y<0$. In the case of $m_y=0$, the situation is beyond Eq. (\ref{Eq27}), that is, $\phi_{L2}=\pi/2$ if $m_x>0$ and $\phi_{L2}=-\pi/2$ if $m_x<0$. In the case of both $m_x=0$ and $m_y=0$, we have $\phi_{L2}=0$. Correspondingly, for those angles in initial state preparation,
we may choose $\theta_{L1}=\arccos(-m_z)$. Due to $\sin(\theta_1)\geq 0$, $\phi_{L1}$ has the same form as $\phi_{L2}$ in Eq. (\ref{Eq27}).

In our experiment, there are six operators required to be measured, i.e., $A_+$, $B_+$, $C_+$, $D_+$, $E_+$ and $F_+$. This is due to fact that our computation of Wasserstein distances encapsulates an added privilege of diminished experimental operations. For example, due to the fact that $|p_{\rho}^{X_{+}}-p_{\rho}^{Y_{+}}| = |p_{\rho}^{X_{-}}-p_{\rho}^{Y_{-}}|$, we are merely required  to carry out experimental manipulations for the determination of former terms $|p_{\rho}^{X_{+}}-p_{\rho}^{Y_{+}}|$. Moreover, $D_+$, $E_+$ and $F_+$ cannot be measured directly by our experimental exertions. So we adopt an alternative approach of joint measurement operator  $M_{\bm{\mu}}$ to figure out the desired values.  For instance, to determine $D_+$, we employ the marginal relation $D_+=\sum_{\mu_2 \mu_3}M_{+\mu_2 \mu_3}$ by explicitly measuring the four operators $M_{+++}$, $M_{++-}$, $M_{+-+}$ and $M_{+--}$.

\section{Experimental implementation}

In the following section we are proceeding to rigorously elucidate the experimental verification of the uncertainty relations perceived in the proceeding section by considering three typical cases concerning the three incompatible observables.

\subsection{Three orthogonal incompatible observables}

\begin{figure*}[hbtp]
\centering
\includegraphics[width=4 cm, height=4.0 cm]{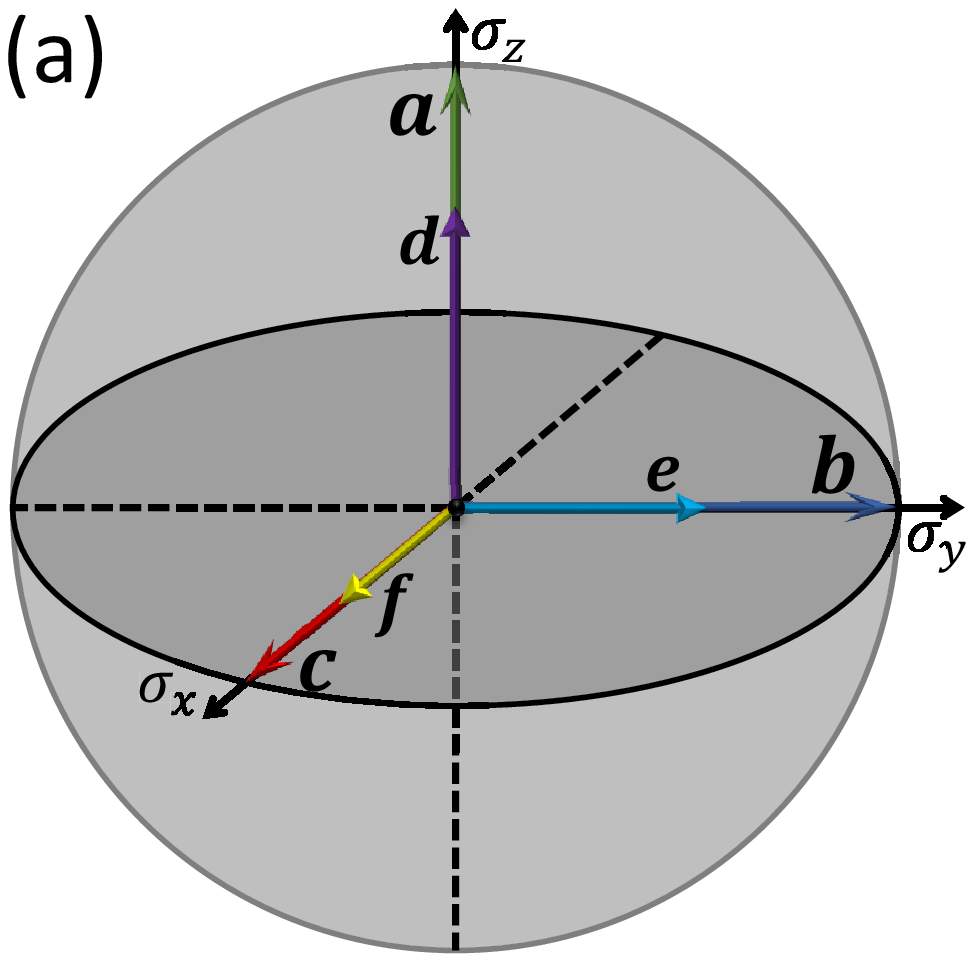}\includegraphics[width=5.0 cm, height=4.0 cm]{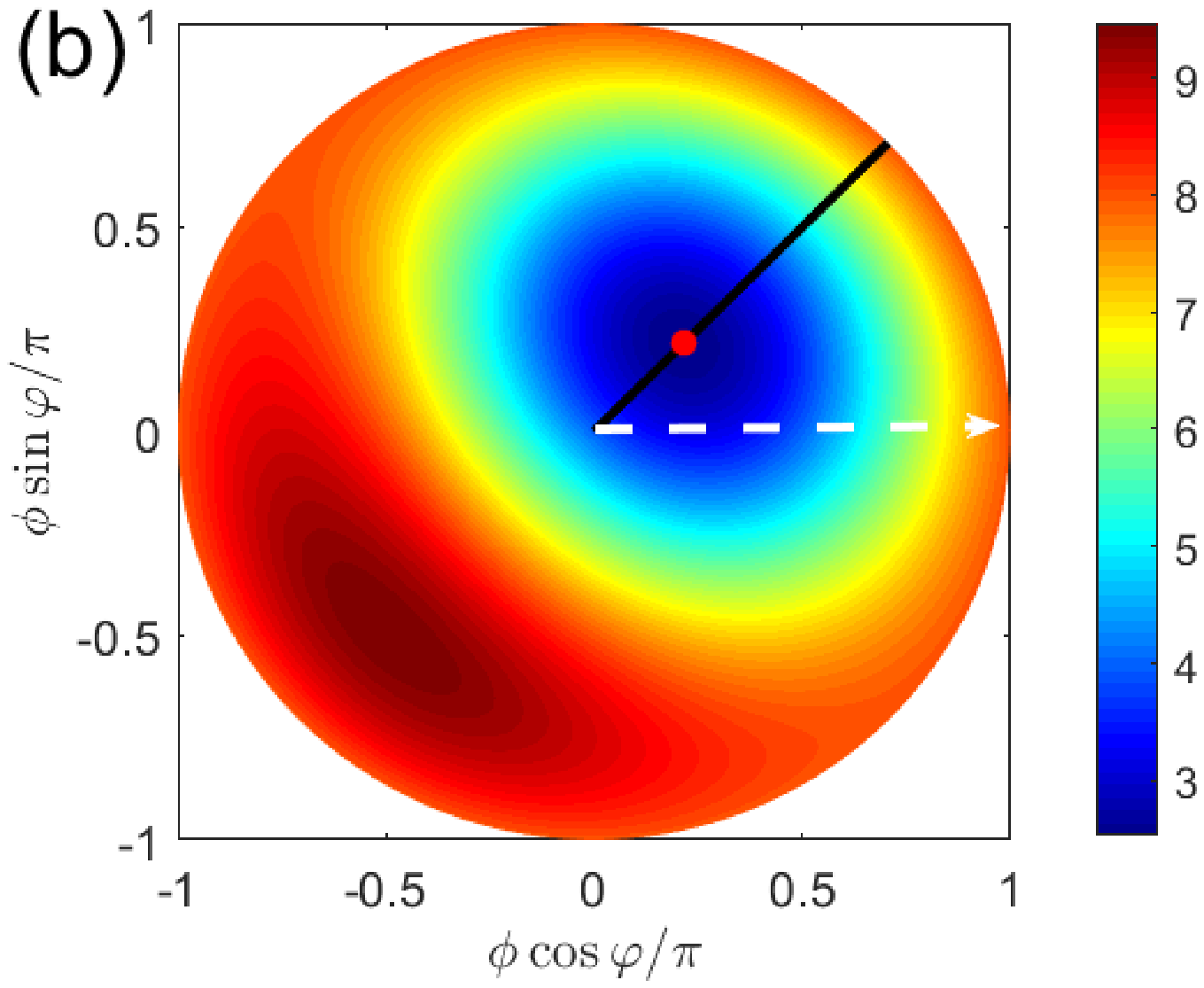}\includegraphics[width=4.5 cm, height=4.0 cm]{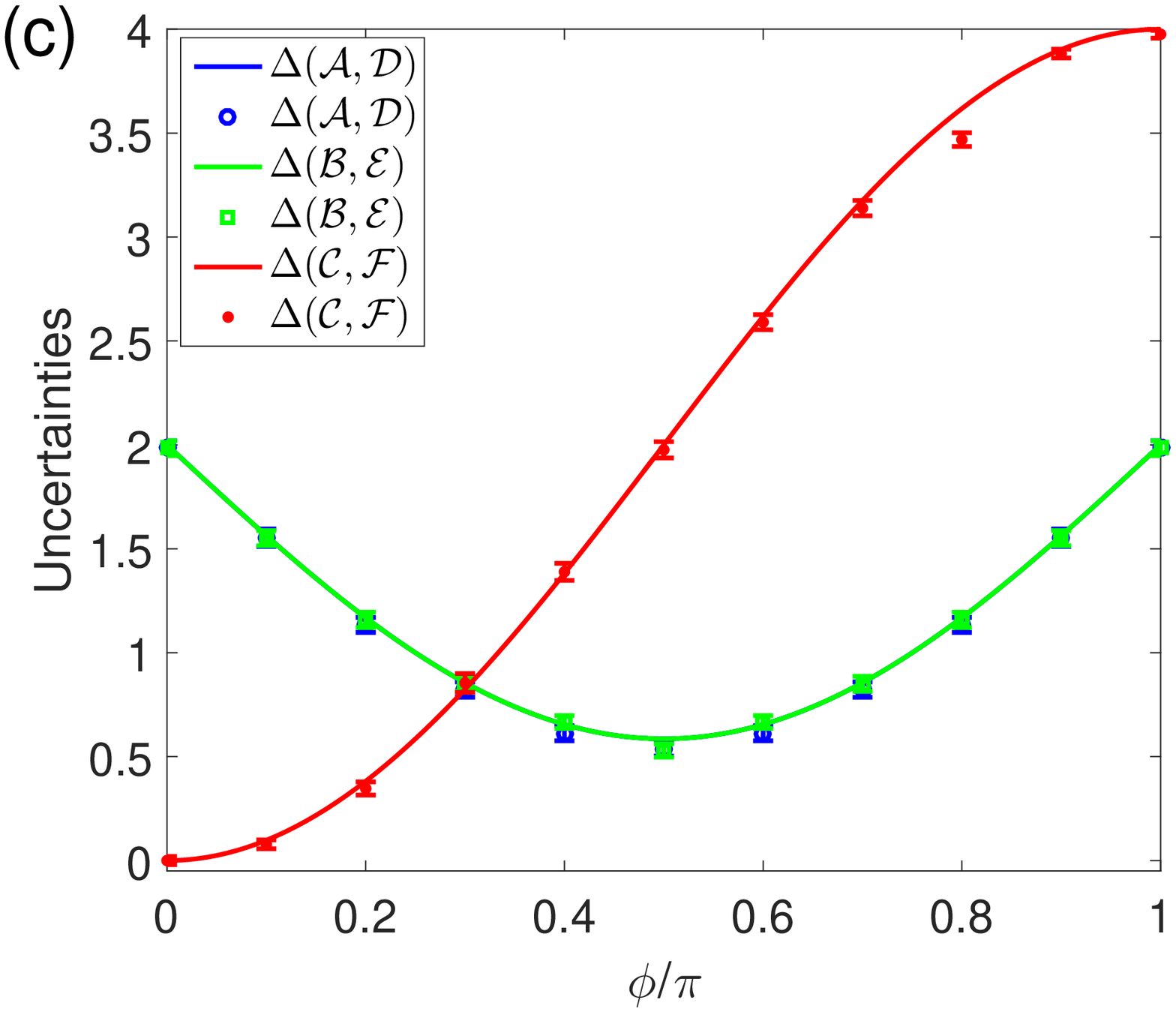}\includegraphics[width=4.5 cm, height=4.0 cm]{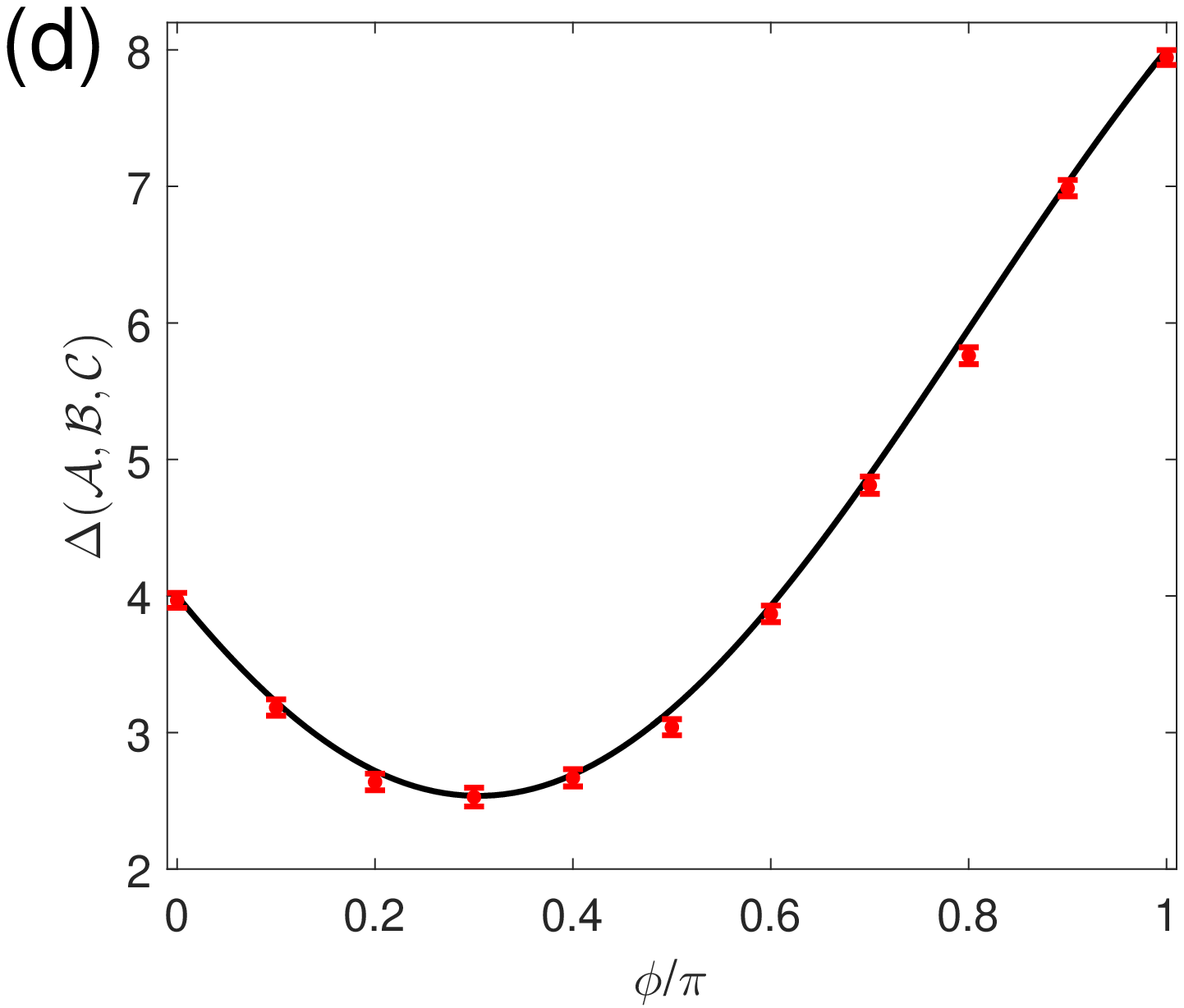}
\caption{(Color online) Three orthogonal incompatible observables $\mathcal{A}, \mathcal{B}, \mathcal{C}$ approximated by three compatible observables  $\mathcal{D}, \mathcal{E}, \mathcal{F}$, respectively. (a) Corresponding vectors of the observables in Bloch sphere. (b) Calculated Heisenberg uncertainty relation $\Delta(\mathcal{A},\mathcal{B},\mathcal{C})$ as functions of the parameters $\phi$ and $\varphi$, corresponding to $k=1$ in the polar coordinate representation with $\phi$ as the radius and $\varphi$ as the polar angle, where the white and black lines denote the polar angle $\varphi=0$ and $\varphi=\pi/4$, respectively, and the red dot denotes the lower bound. (c) Experimental measurements of the separate terms $\Delta(\mathcal{A},\mathcal{D})$, $\Delta(\mathcal{B},\mathcal{E})$ and $\Delta(\mathcal{C},\mathcal{F})$ of the Heisenberg uncertainty relation, where the curves are the analytical results, and the blue and green data are nearly overlapped due to the very close values in between. (d) Heisenberg uncertainty relation for three orthogonal incompatible observables $\mathcal{A}$, $\mathcal{B}$ and $\mathcal{C}$ with the dots and curve denoting the experimental data and analytical result, respectively. Error bars in (c) and (d) indicate the standard deviation of the data with each measured with repetition of 20,000 times. }
\label{Fig4}
\end{figure*}

\begin{table*}
\caption{Parameter values for the measurement pulses in observing $A_{+}$, $B_{+}$, $C_+$ and the joint measurement operator $M_{\mu_1 \mu_2 \mu_3}$ for three orthogonal incompatible observables $\mathcal{A}=\sigma_z$, $\mathcal{B}=\sigma_y$ and $\mathcal{C}=\sigma_x$. We set $\tilde{\theta}_2=\arccos(\sqrt{2}\sin\varphi/2)$, $\bar{\theta}_2=\arccos(-\sqrt{2}\sin\varphi/2)$  and $\bar{\phi}=\arctan\sqrt{2}\cot\varphi$.}
\centering
\begin{tabular}{cccccccccccccccccccccccccccc}
\hline
\hline
  &$A_{+}$ & $B_{+}$ & $C_{+}$ &$M_{+++}$&$M_{++-}$&$M_{+-+}$&$M_{+--}$&$M_{-++}$&$M_{-+-}$&$M_{--+}$&$M_{---}$ \\
  \hline
$\theta_{L2}$ &0 &$\pi/2$&$\pi/2$ &$\tilde{\theta}_2$&$\tilde{\theta}_2$&$\tilde{\theta}_2$&$\tilde{\theta}_2$&$\bar{\theta}_2$ &$\bar{\theta}_2$ &$\bar{\theta}_2$ &$\bar{\theta}_2$   \\
$\phi_{L2}$ & 0 &  0& $\pi/2$&$\bar{\phi}$&$-\bar{\phi}$&$\pi-\bar{\phi}$&$\pi+\bar{\phi}$ &$\bar{\phi}$&$-\bar{\phi}$&$\pi-\bar{\phi}$&$\pi+\bar{\phi}$\\
\hline
\hline
\end{tabular}
\label{Table1}
\end{table*}

In this section,  we proceed to demonstrate the case for three incompatible observables when they are oriented orthogonally. As presented in Fig. \ref{Fig4}(a)  the mutually perpendicular incompatible observables can be given as $\mathcal{A}=\sigma_z$, $\mathcal{B}=\sigma_y$ and $\mathcal{C}=\sigma_x$ such that the corresponding $\bm{a},~\bm{b}$ and $\bm{c}$ are $ \bm(0,0,1),\bm(0,1,0)$ and $(1,0,0) $ respectively. Following the steps as in Appendix B, we can write the approximations $ \bm{d},\bm{e} $ and $\bm{f} $ as
\begin{eqnarray}
&&\bm{d}=(0,0,k\sin\varphi\sin\phi),\ \bm{e}=(0,k\cos\varphi\sin\phi,0), \notag \\
&&\qquad \  \bm{f}=(k\cos\phi,0,0),
\label{Eq31}
\end{eqnarray}
Thus the uncertainty relation for these three observables is given by
\begin{equation}
\Delta(\mathcal{A},\mathcal{B},\mathcal{C})\geq 2\sqrt{3}(\sqrt{3}-1),
\label{Eqs29}
\end{equation}
where the lower bound appears at $k=1$, $\varphi=\pi/4$ and $\phi=\arccos\sqrt{1/3}$, as explained in Appendix B.

Keeping in view the constraints in a single qubit-system, we need to scrutinize the possibility for the predominating boundry conditions. The optimal approximations are always obtained at the boundary of Eq. (\ref{Eq3}) with $k=1$. Utilizing Eq. (\ref{Eq4}), we can obtain the joint measurement operators for $\bm{d}$, $\bm{e}$ and $\bm{f}$, from which one can readily obtain $G=1+\sum_{j>j}\mu_i\mu_j\bm{\lambda}_i\cdot\bm{\lambda}_j=1$. Considering $\text{Rank}[M_{\mu}]=1$ as the paramount condition referring to the measurement prospects of $M_{\bm{\mu}}$ in one-qubit system  as detailed in \cite{SA-2-e1600578}, we have,
\begin{equation}
G^2=\sum_{j=1,2,3}(\sum_i\mu_i\lambda_i^j)^2,
\label{Eq30}
\end{equation}
where $\lambda_i^j$ denotes the $j$th element of the vector $\bm{\lambda}_i$. Additionally from Eq. (\ref{Eqs29}), we know $\sum_{j=1,2,3}(\sum_i\mu_i\lambda_i^j)^2=k^2$. Considering Eq. (\ref{Eq30}) and using the result $G=1$, we obtain $k=1$, which is consistent with the lower bound condition in Eq. (\ref{Eqs29}), implying that the lower bound in this case could be reached by the one-qubit measurements.

In our experiment we illustrate the case related to $\varphi=\pi/4$, where the lower bound of the uncertainty relation exists along the black line as plotted in Fig. \ref{Fig4}(b).
From Eq. (\ref{Eq8}) and Eq. (\ref{Eqs29}), we can easily find that the optimal states for $\Delta_{\rho}(\mathcal{A},\mathcal{D})$, $\Delta_{\rho}(\mathcal{B},\mathcal{E})$ and $\Delta_{\rho}(\mathcal{C},\mathcal{F})$ are $\rho_1=(1+\sigma_z)/2$, $\rho_2=(1+\sigma_y)/2$ and $\rho_3=(1+\sigma_x)/2$, respectively. In this case, we have analytical results for the experimentally measurable quantities, i.e., $p_{\rho_1}^{K_{+}}=1$ with $K=A,B,C$, $p_{\rho_1}^{M_{+\mu_2 \mu_3}}=(1+k\sin\varphi\sin\phi)/8$, $p_{\rho_2}^{M_{\mu_1 + \mu_3}}=(1+k\cos\varphi\sin\phi)/8$, and $p_{\rho_3}^{M_{\mu_1 \mu_2 +}}=(1+k\cos\phi)/8$ for all the possibilities of $\mu_1,\mu_2,\mu_3=\pm$. In particular, for the given angle $\varphi=\pi/4$ (i.e., the black line in Fig. \ref{Fig4}(b)), we have $p_{\rho_1}^{M_{+\mu_2 \mu_3}}=p_{\rho_2}^{M_{\mu_1 +\mu_3}}$.

With the laser irradiation as set by the parameter values as listed in Table \ref{Table1}, we have experimentally measured the uncertainty relations.
Fig. \ref{Fig4}(c) demonstrates that the separate terms of Eq. (\ref{Eq7}), i.e., $\Delta(\mathcal{A},\mathcal{D})$ and $\Delta(\mathcal{B},\mathcal{E})$ behave nearly the same whereas $\Delta(\mathcal{C},\mathcal{F})$ varies very differently. This observation is resulted from the spatial asymmetry of $\bm{f}$ with respect to $\bm{d}$ and $\bm{e}$. Nevertheless, the three curves have a point of intersection at $\phi=\arccos\sqrt{1/3}$, at which the three approximate observables have the corresponding vectors of the same length $1/\sqrt{3}$ and deviate from their target incompatible observables with the same uncertainty. Moreover, as plotted in Fig. \ref{Fig4}(d), the lower bound also occurs at $\phi=\arccos\sqrt{1/3}$, which agrees with the prediction of Eq. (\ref{Eqs29}).


\subsection{Three coplanar incompatible observables}

\begin{figure*}[hbtp]
\centering
\includegraphics[width=5.5 cm, height=4.5 cm]{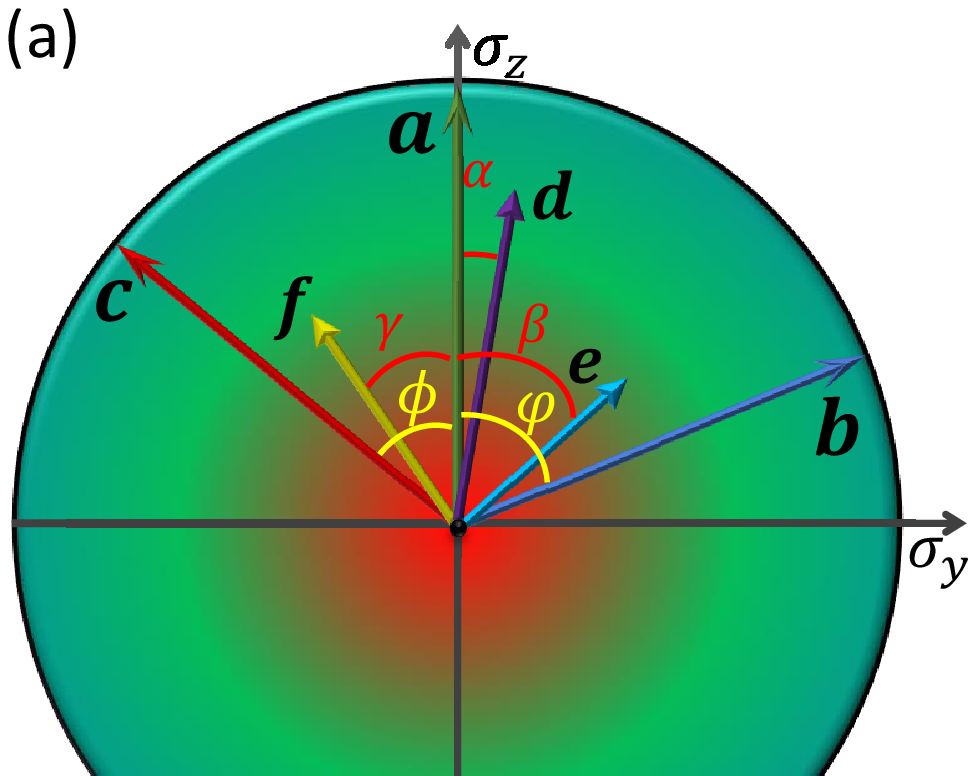}
\includegraphics[width=6.5 cm, height=4.5 cm]{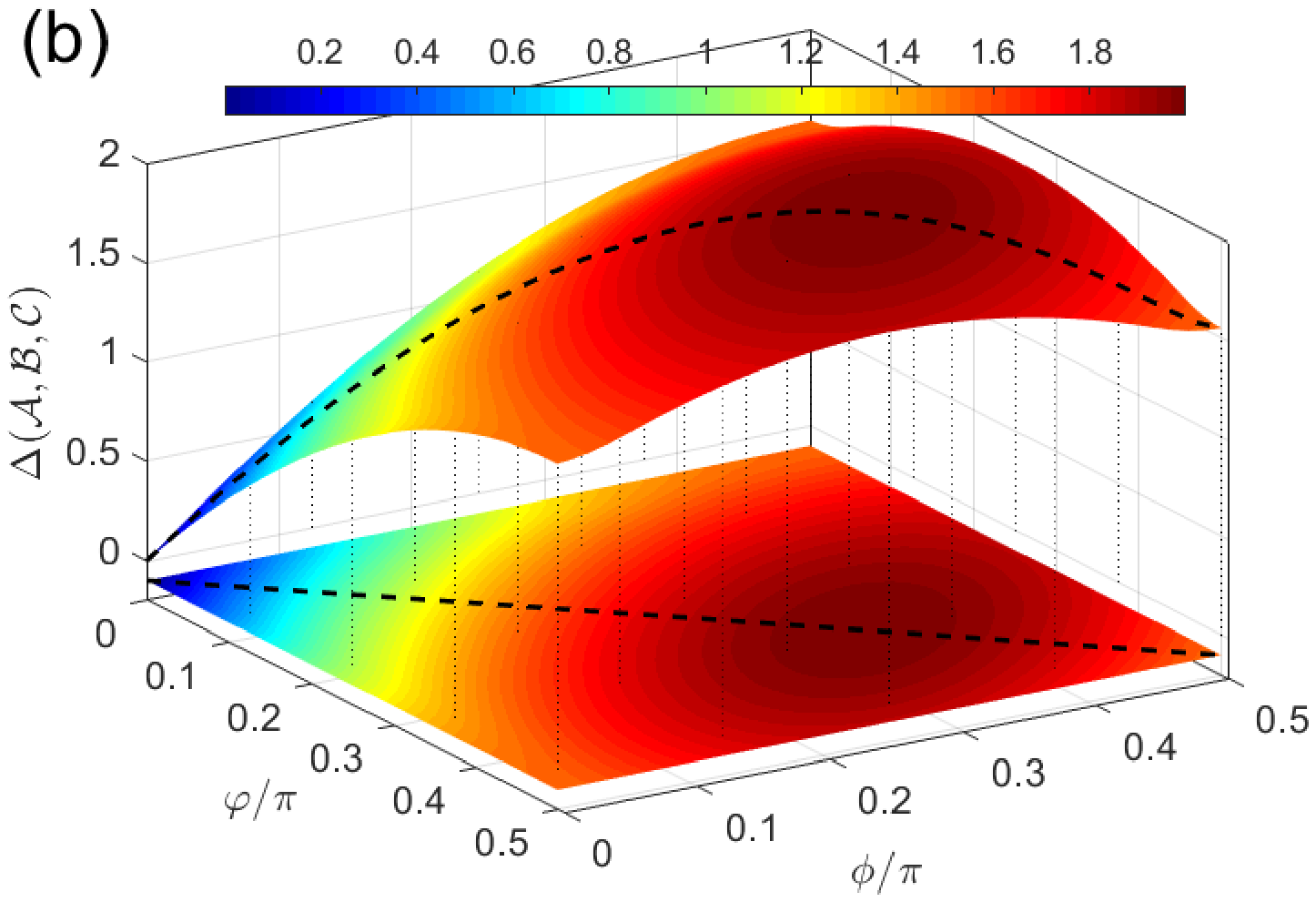}
\includegraphics[width=5.5 cm, height=4.5 cm]{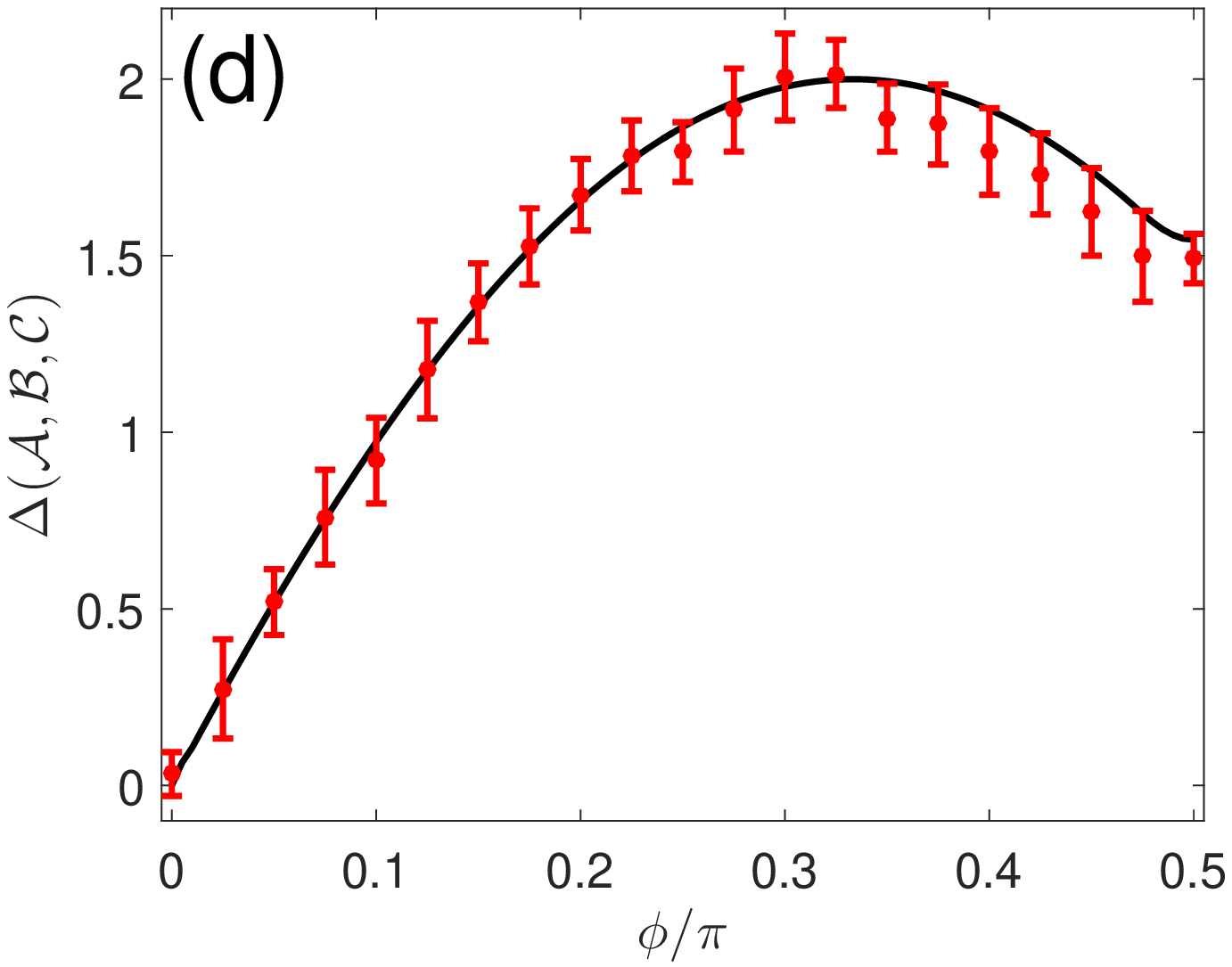}
\includegraphics[width=18.0 cm, height=10.0 cm]{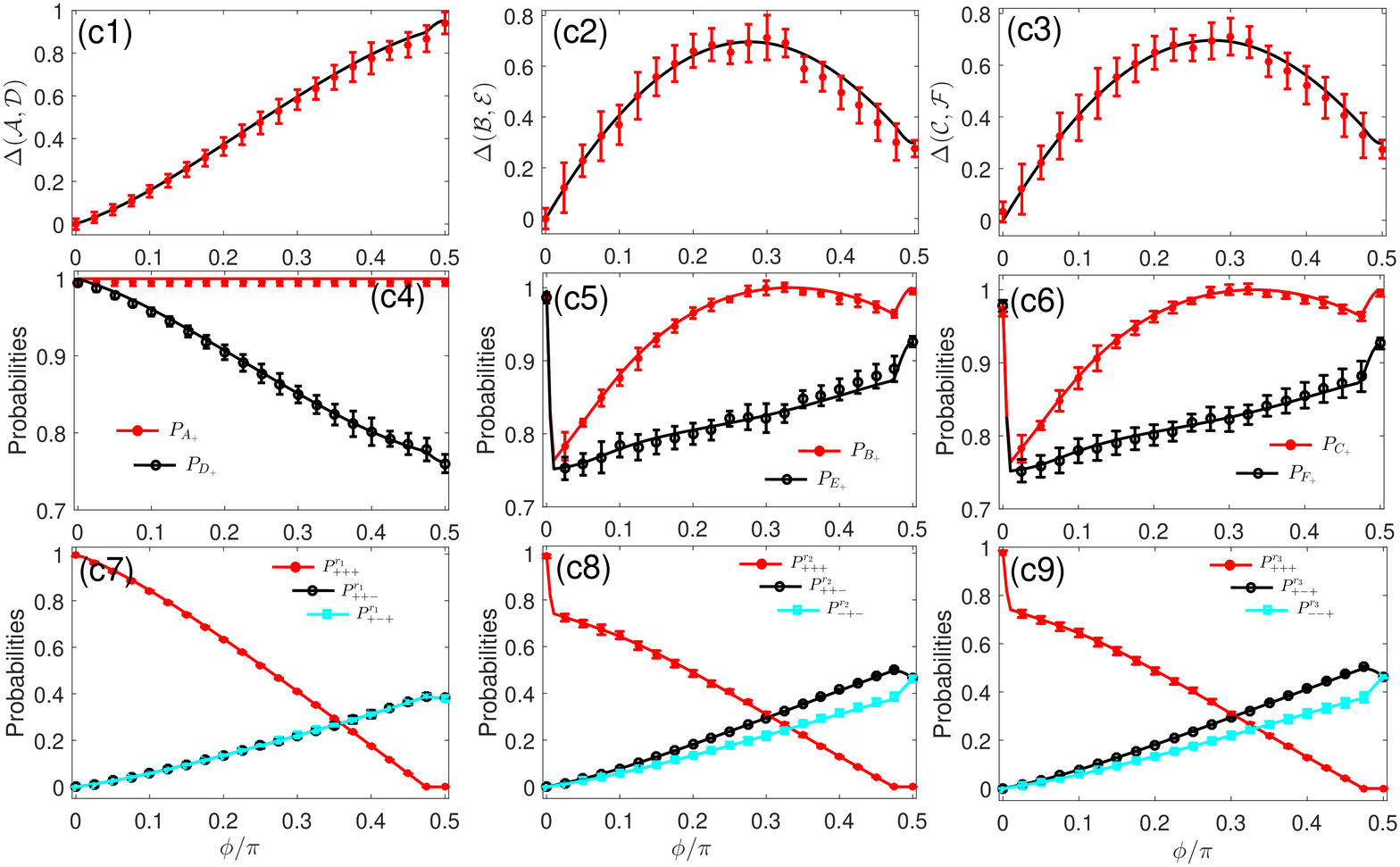}
\caption{(Color online) Three coplanar incompatible observables $\mathcal{A}, \mathcal{B}, \mathcal{C}$ approximated by three compatible observables $\mathcal{D}, \mathcal{E}, \mathcal{F}$, respectively. (a) Corresponding vectors of the observables in Bloch sphere, where $\alpha$, $\beta$, $\gamma$ are, respectively, angles of the vectors $\bm{d}$, $\bm{e}$, $\bm{f}$ with respect to $\sigma_{z}$. (b) Calculated Heisenberg uncertainty relation $\Delta(\mathcal{A},\mathcal{B},\mathcal{C})$ as functions of $\phi$ and $\varphi$, where the black dashed lines denote $\phi=\varphi$. (c1-c3) Experimental measurements of  the separate terms $\Delta(\mathcal{A},\mathcal{D})$, $\Delta(\mathcal{B},\mathcal{E})$ and $\Delta(\mathcal{C},\mathcal{F})$ of the Heisenberg uncertainty relation, where the solid curves are numerical results. (c4-c6) Experimental measurements of the positive operators $A_+,B_+,C_+$ (red dots) and $D_+,E_+,F_+$ (black circles). (c7-c9) Experimental measurements of the joint measurement operators $M_{\mu_1 \mu_2 \mu_3}$ for $D_+$ (c7), $E_+$ (c8) and $F_+$ (c9), where $P^{r_k}_{\mu_1 \mu_2 \mu_3}=p_{\rho_k}^{M_{\mu_1 \mu_2 \mu_3}}$, and no $P^{r_1}_{+--}$, $P^{r_2}_{-++}$ or $P^{r_3}_{-+-}$ exists.  All the solid curves in (c1-c9) are from numerical treatments. (d) Experimental observation of the Heisenberg uncertainty relation for the case $\phi=\varphi$ with the dots and curve denoting the experimental values and numerical result, respectively. Error bars in (c) and (d) indicate the standard deviation of the data with each measured with repetition of 20,000 times.}
\label{Fig5}
\end{figure*}

\begin{table}
\caption{Parameter values for the measurement pulses in observing $A_{+}$, $B_{+}$, $C_+$ for three coplanar incompatible observables given in Eq. (\ref{Eq31}). The values are calculated numerically due to no analytical result for $M_{\mu_1 \mu_2 \mu_3}$.}
\centering
\begin{tabular}{cccccccccccccccccccccccccccc}
\hline
\hline
  &$A_{+}$ & $B_{+}$ & $C_{+}$  \\
  \hline
$\theta_{L2}$ &0 &$\varphi$&$\phi$   \\
$\phi_{L2}$ & 0 &  0& $\pi$\\
\hline
\hline
\end{tabular}
\label{Table2}
\end{table}

In this section, we consider three coplanar incompatible observables generally described by,
\begin{equation}
\bm{x}=(0,\sin\theta_x,\cos\theta_x),
\label{Eq18}
\end{equation}
with $x=a,b, c$ satisfying $|\bm{x}|=1$. By virtue of unitary transformation, any set of three coplanar observables in the Bloch sphere can be transformed to attain the above representation.
In contrast to the above subsection, no analytical solution can be found in this case to compute the lower bound. As a result, here we implement a different way of finding the solution for the optimal function $\bar{\Delta}_{lb2}$ associated with the lower bound, based on Eq. (\ref{Eq21}) in Appendix C. So the uncertainty relation is given by
\begin{equation}
\Delta(\mathcal{A},\mathcal{B},\mathcal{C})\geq \bar{\Delta}_{lb2}.
\end{equation}
In our experiment, we exemplify the case for which $\mathcal{A}=\sigma_z$, $\mathcal{B}=\sin\varphi\sigma_y+\cos\varphi\sigma_z$ and $\mathcal{C}=\sin\phi\sigma_y+\cos\phi\sigma_z$, implying
\begin{eqnarray}
&&\bm{a}=(0,0,1),\ \bm{b}=(0,\sin\varphi,\cos\varphi), \notag \\
&&\qquad \  \bm{c}=(0,-\sin\phi,\cos\phi),
\label{Eq31}
\end{eqnarray}
as sketched in Fig. \ref{Fig5}(a). Since no analytical solution is available for the associated vectors to approximate the observables $\mathcal{D}$, $\mathcal{E}$ and $\mathcal{F}$, we implement a numerical solution based on Eq. (\ref{Eq21}), by which the precise approximation to the incompatible observables $\mathcal{A}$, $\mathcal{B}$ and $\mathcal{C}$ are gained by explicit adjustment of the parameters $\varphi$ and $\phi$.

Fig. \ref{Fig5}(b) indicates that the uncertainty increases with $\varphi$ and $\phi$, where the maximum uncertainty appears along the diagonal line $\phi=\varphi$. Numerical result shows that the maximum uncertainty appears in the case of $\bar{\Delta}_{lb2}=2$ at the position $\phi=\varphi=\pi/3$, implying that $ \Delta(\mathcal{A},\mathcal{B},\mathcal{C})\geq 2$ with $\mathcal{A}=\sigma_z$, $\mathcal{B}=\sigma_z/2+\sqrt{2}\sigma_y/\sqrt{3}$ and $\mathcal{C}=\sigma_z/2-\sqrt{2}\sigma_y/\sqrt{3}$. This indicates that $\bm{a}=\bm{b}+\bm{c}$, forming a regular triangle. In fact, at this point we also have the relation $\Delta(\mathcal{A},\mathcal{D})=\Delta(\mathcal{B},\mathcal{E})=\Delta(\mathcal{C},\mathcal{F})=2/3$. In addition, we mention that validity of the coplanar condition for $\bm{d}$, $\bm{e}$ and $ \bm{f}$ has been checked in the numerical calculation for Fig. \ref{Fig5}(b), which shows the module $|\bm{d}\times\bm{e}\cdot\bm{f}|<10^{-5}$.

Our experimental measurement takes the case regarding the diagonal line $\phi=\varphi$ as an example to verify the uncertainty relation. The data sets of the probabilities regarding the observables have been measured, as presented in the panels of Fig. \ref{Fig5}(c). We find that, due to the identical change of $\phi$ and $\varphi$, $\Delta(\mathcal{B},\mathcal{E})$ and $\Delta(\mathcal{C},\mathcal{F})$ behave nearly the same as parabolas whereas $\Delta(\mathcal{A},\mathcal{D})$ varies differently as a monotonous increase with $\phi$. The three curves have the common point at $\phi=0$ which means a trivial solution since the three curves are collinear and zero uncertainty exists in this case. The three curves intersect at $\phi=\pi/3$, which, from the state-independent Heisenberg uncertainty relation plotted in Fig. \ref{Fig5}(d), is the point corresponding to the maximum of the lower bound.

The experimental results plotted in Fig. \ref{Fig5}(c) confirm the numerical prediction for the maximum uncertainty. Since we suppose that $\bm{b}$ and $\bm{c}$ are positioned in opposite sides of $\bm{a}$, we find the optimal approximation of the vector $\bm{d}$ to the target vector $\bm{a}$ always existing along the direction of $\bm{a}$. As such, we may only consider the case of $\alpha=0$ in Fig. \ref{Fig5}(a). In this case, when $\phi=0$ turns to be $\phi\neq$0, the measurement values of $A_+$ keep constant and $D_+$ decreases slightly, whereas for other observables, including $B_+$, $C_+$, $E_+$ and $F_+$, the measured possibilities drop largely, see shown in Fig. \ref{Fig5}(c4-c6). The latter is due to the fact that the optimal vector representing $E_+$ ($F_+$) should have a large deviation from the vector $\bm{b}$ ($\bm{c}$), which actually reflects the restriction imposed by the uncertainty relation. Nevertheless, as the dropping occurs in $B_+$ ($C_+$) and $E_+$ ($F_+$) nearly identically, no discontinuity can be found in the uncertainty relations $\Delta(\mathcal{B},\mathcal{E})$ and $\Delta(\mathcal{C},\mathcal{F})$ in Fig. \ref{Fig5}(c2,c3). As such, we have a complete and continuous observation of the Heisenberg uncertainty relation in Fig. \ref{Fig5}(d).


\subsection{Three incompatible observables with one observable orthogonal to the other two}

\begin{figure*}[hbtp]
\centering
\includegraphics[width=5.5 cm, height=4.8 cm]{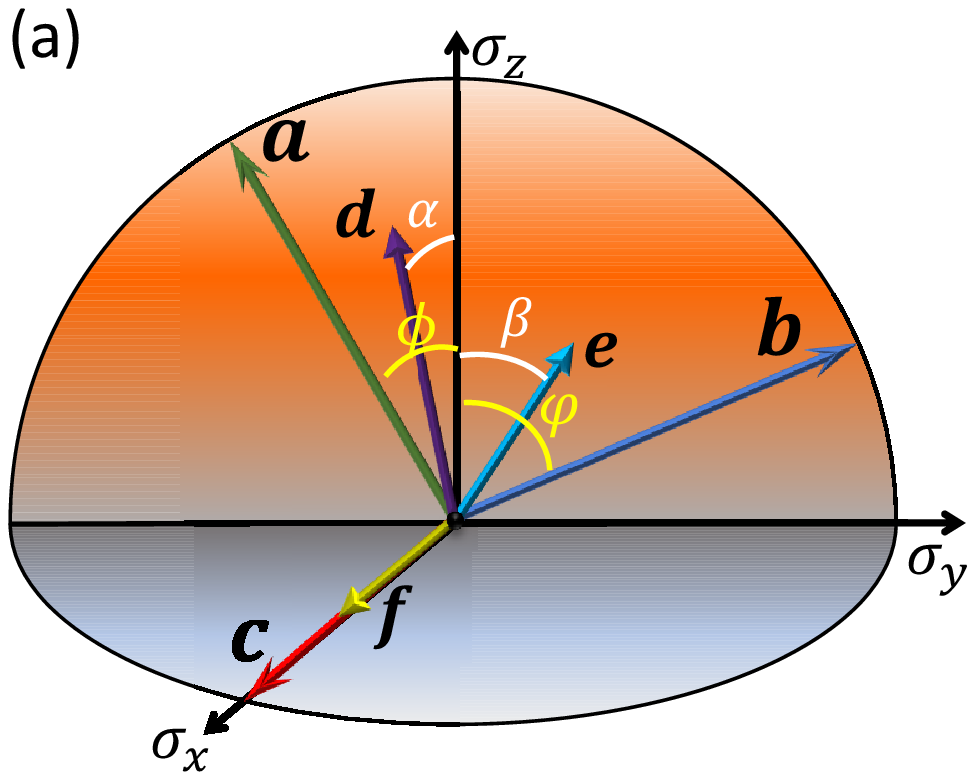}
\includegraphics[width=6 cm, height=4.8 cm]{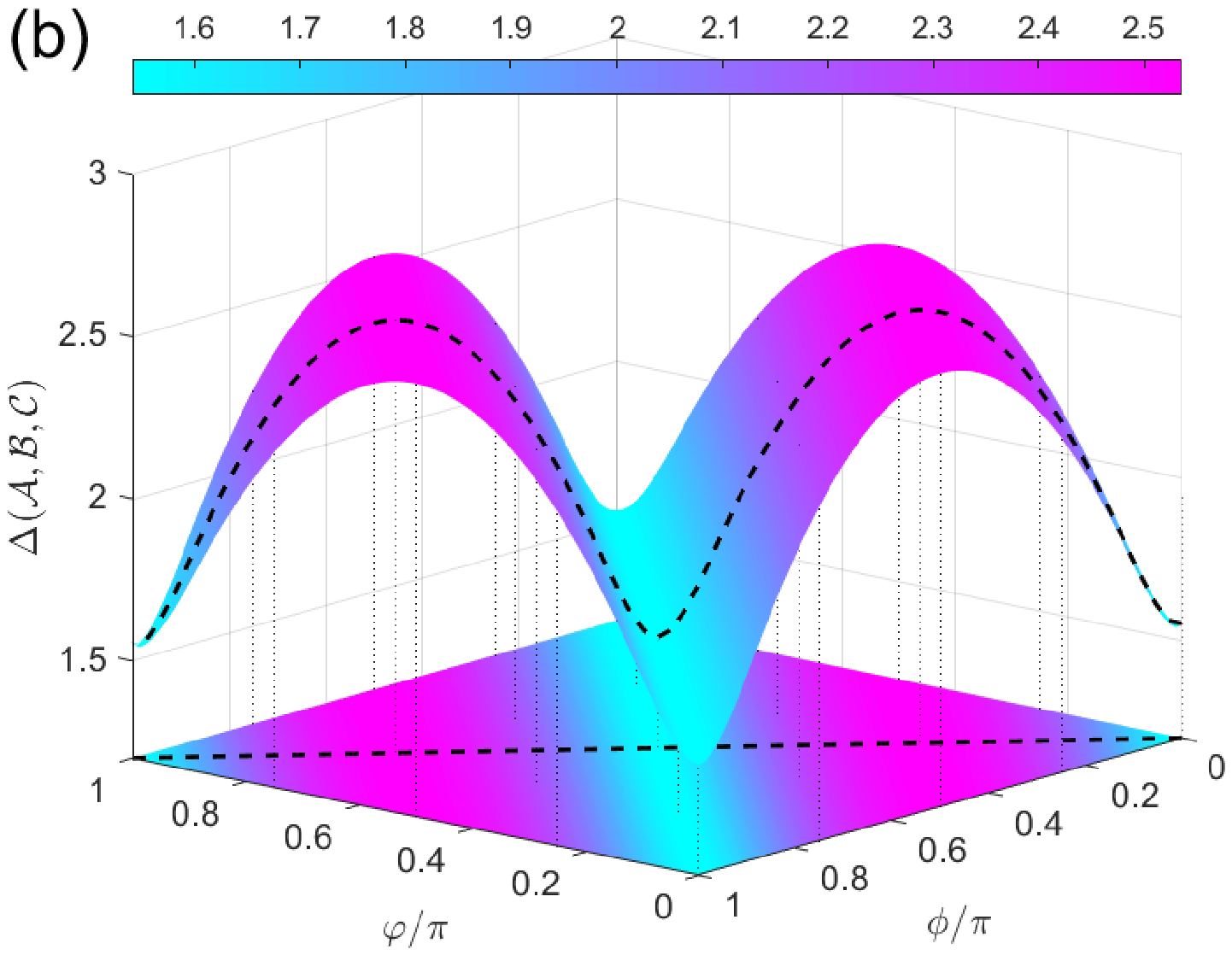}
\includegraphics[width=6 cm, height=4.8 cm]{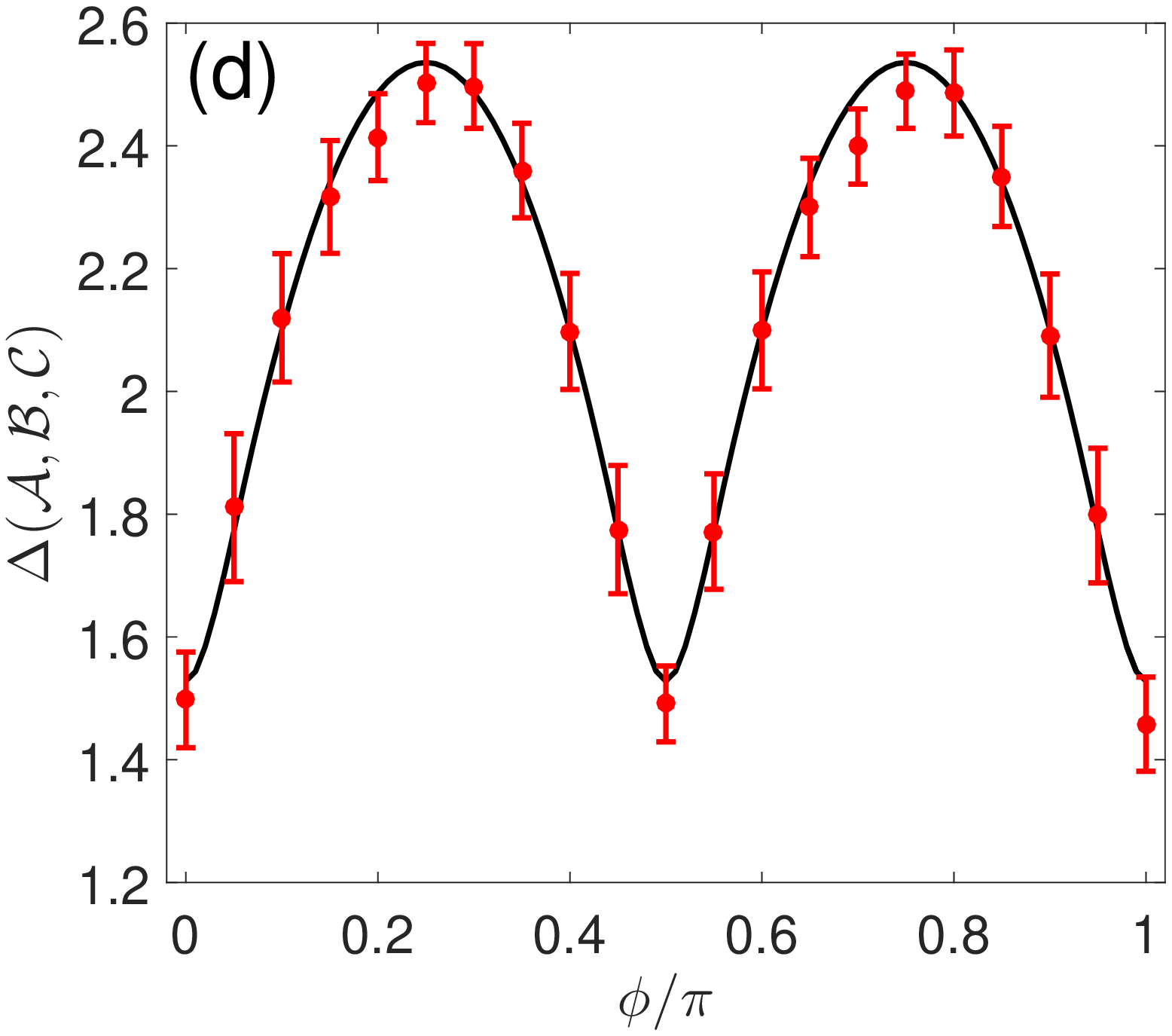}
\includegraphics[width=18 cm, height=10 cm]{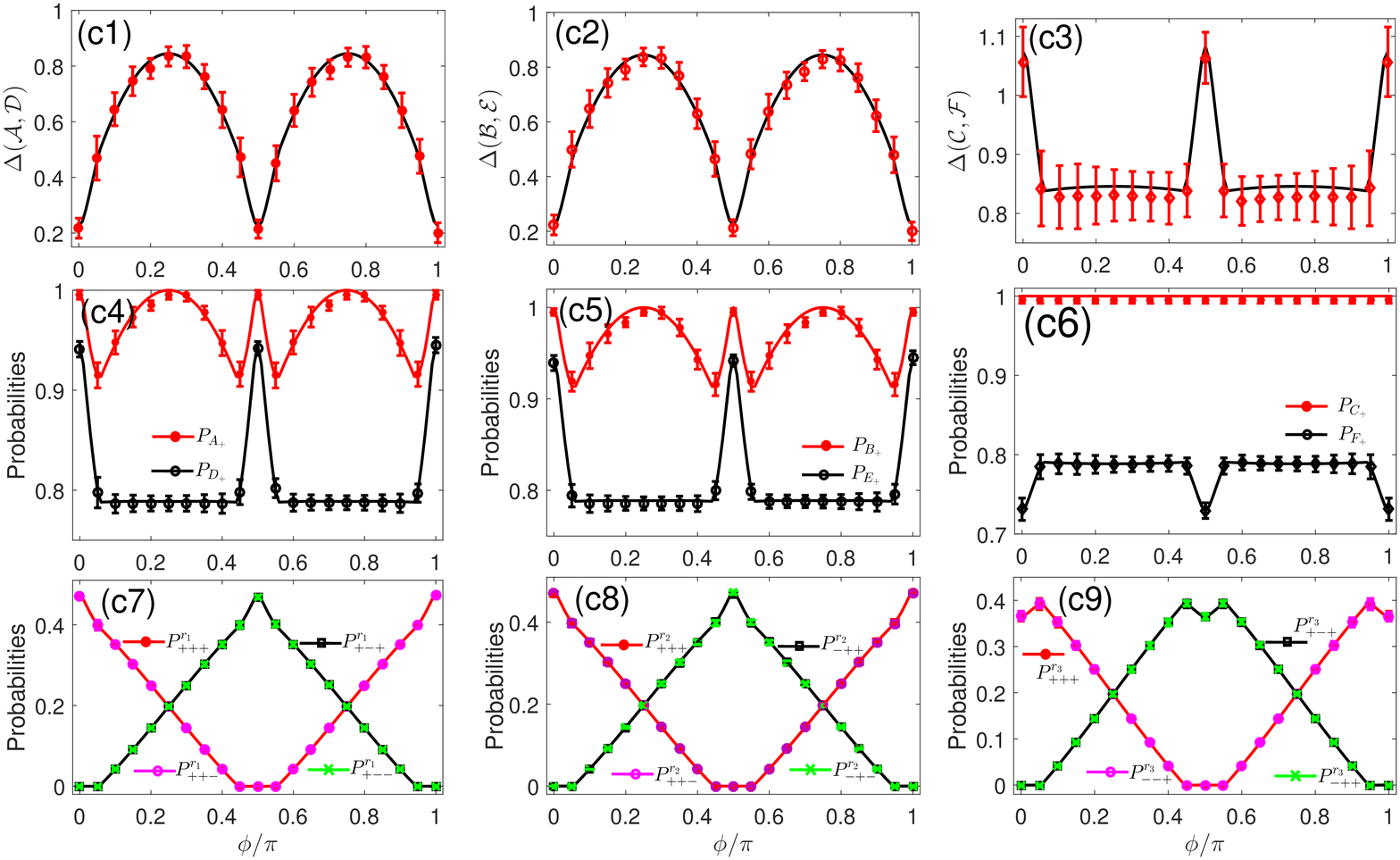}
\caption{(Color online) Three incompatible observables $\mathcal{A}, \mathcal{B}, \mathcal{C}$ in the case of $\bm{c}\perp \bm{a},\bm{b}$ approximated by three compatible observables $\mathcal{D}, \mathcal{E}, \mathcal{F}$, respectively. (a) Corresponding vectors of the observables in Bloch sphere, where $\alpha$ and $\beta$ are, respectively, angles of the vectors $\bm{d}$ and $\bm{e}$ with respect to $\sigma_{z}$. (b) Calculated Heisenberg uncertainty relation $\Delta(\mathcal{A},\mathcal{B},\mathcal{C})$ as functions of $\phi$ and $\varphi$, where the black dashed lines denote $\phi=\varphi$. (c1-c3) Experimental measurements of the separate terms $\Delta(\mathcal{A},\mathcal{D})$, $\Delta(\mathcal{B},\mathcal{E})$ and $\Delta(\mathcal{C},\mathcal{F})$ of the Heisenberg uncertainty relation, where the solid curves are numerical results. (c4-c6) Experimental measurements of the operators $P_{A_+,B_+,C_+}$ (red dots) and $P_{D_+,E_+,F_+}$ (black circles), where the solid curves are numerical results. (c7-c9) Experimental measurements of the joint measurement operators $M_{\mu_1 \mu_2 \mu_3}$ for $D_+$ (c7), $E_+$ (c8) and $F_+$ (c9), where $P^{r_k}_{\mu_1 \mu_2 \mu_3}=p_{\rho_k}^{M_{\mu_1 \mu_2 \mu_3}}$. In (c7) and (c8), the values have the relations $P^{r_1}_{M_{+++}}=P^{r_1}_{M_{++-}}$, $P^{r_1}_{M_{+-+}}=P^{r_1}_{M_{+--}}$, $P^{r_2}_{M_{+++}}=P^{r_2}_{M_{++-}}$ and $P^{r_2}_{M_{-++}}=P^{r_2}_{M_{-+-}}$. In (c9), the values have the relation $P^{r_3}_{M_{+++}}=P^{r_3}_{M_{--+}}$ and $P^{r_3}_{M_{+-+}}=P^{r_3}_{M_{-++}}$. (d) Experimental observation of the Heisenberg uncertainty relation in the case of $\phi=\varphi$ with the dots and curve denoting the experimental and numerical results, respectively. Error bars in (c) and (d) indicate the standard deviation of the data with each measured with repetition of 20,000 times.}
\label{Fig6}
\end{figure*}

\begin{table}
\caption{Parameter values for the measurement pulses in observing $A_{+}$, $B_{+}$, $C_+$ for $\bm{c}\perp \bm{a},\bm{b}$ as designed in Eq. (\ref{Eq32}). The values are calculated numerically due to no analytical solution.}
\centering
\begin{tabular}{cccccccccccccccccccccccccccc}
\hline
\hline
  &$A_{+}$ & $B_{+}$ & $C_{+}$  \\
  \hline
$\theta_{L2}$ &$\phi$ &$\varphi$&$\pi/2$   \\
$\phi_{L2}$ & $\pi$ &  0& $\pi/2$\\
\hline
\hline
\end{tabular}
\label{Table3}
\end{table}

This subsection deals with the three incompatible observables with one of them orthogonal to the other two, whose substantial representation is as follows,
\begin{equation}
\bm{x}=(\sin\theta_x,\cos\theta_x,0),\ \bm{c}=(1,0,0),
\label{Eq22}
\end{equation}
with $x=a,b$ satisfying $|\bm{x}|=1$. The uncertainty relation in this case is assumed to be
\begin{equation}
\Delta(\mathcal{A},\mathcal{B},\mathcal{C})\geq \bar{\Delta}_{lb3},
\end{equation}
where $\bar{\Delta}_{lb3}$ is the optimal function associated with the lower bound and can be numerically solved by Eq. (\ref{Eq24}) in Appendix D. For the sake of convenience of the experimental demonstration, we consider, as sketched in Fig. \ref{Fig6}(a), the three incompatible observables with the following form
 \begin{eqnarray}
&&\bm{a}=(0,-\sin\phi,\cos\phi),\ \bm{b}=(0,\sin\varphi,\cos\varphi), \notag \\
&&\qquad\qquad \qquad \bm{c}=(1,0,0),
\label{Eq32}
\end{eqnarray}
where $\phi$ ($\varphi$) denotes the intersection angle of $\bm{a}$ ($\bm{b}$) with respect to $\sigma_z$ direction.

As shown in Fig. \ref{Fig6}(b), the uncertainty varies with $\varphi$ and $\phi$, reaching both the maximum and minimum along the diagonal line $\phi=\varphi$.
This is due to the symmetry of the vectors $\bm{a}$ and $\bm{b}$ in $\sigma_y-\sigma_z$ plane. As such, we have two maxima of the uncertainty, i.e., $\bar{\Delta}_{lb3}^{\max}=2\sqrt{3}(\sqrt{3}-1)$ at $\phi=\varphi=\pi/4$ and $3\pi/4$, which actually corresponds to the situation where all of the three observables are in mutual orthogonal orientation. The minima of the uncertainty appear at $\phi=\varphi=0$ and $\pi/2$, where the vectors $\bm{a}$ and $\bm{b}$ are collinear and $\bar{\Delta}_{lb3}^{\min}=1.55$. We also have to mention that we have found in the calculation for Fig. \ref{Fig6}(b) the module $|\bm{d}\cdot\bm{f}|+|\bm{e}\cdot\bm{f}|<10^{-4}$, implying $\bm{f}\perp \bm{d},\bm{e}$ is satisfied.

Our experimental measurements in this case exemplify the case of the diagonal line $\phi=\varphi$. The measured data sets for the observables' probabilities in Fig. \ref{Fig6}(c) demonstrate that all the three terms of the uncertainty relation vary in symmetry with respect to $\phi=\pi/2$, where $\Delta(\mathcal{A},\mathcal{D})$ and $\Delta(\mathcal{B},\mathcal{E})$ behave nearly the same whereas $\Delta(\mathcal{C},\mathcal{F})$ varies very differently. This is due to the symmetry of $\bm{a}$ and $\bm{b}$ in $\sigma_y-\sigma_z$ plane and the orthogonality of $\bm{c}$ with $\bm{a},\bm{b}$. 
We find some peculiar characteristics in Fig. \ref{Fig6}(c1-c6). In Fig. \ref{Fig6}(c4,c5), $P_{K_+}$ with $K=D,E$ keep almost constant for most values of $\phi$, while jumping to the maximum around $\phi/\pi=1/2$ and 1.  Correspondingly, both $\Delta(\mathcal{A},\mathcal{D})$ and $\Delta(\mathcal{B},\mathcal{E})$ in Fig. \ref{Fig6}(c1,c2) behave like the absolute value of sine function with respect to  $\phi=\pi/2$. In contrast, $\Delta(\mathcal{C},\mathcal{F})$ exhibits a different behavior, but with similarity to $P_{F_+}$. This observation reflects the fact, as displayed in Fig. \ref{Fig6}(a), that $\bm{f}$ is always along the direction of $\bm{c}$, whereas $\bm{d}$ and $\bm{e}$ have variational intersection angles with $\bm{a}$ and $\bm{b}$, respectively.
With combination of the results in Fig. \ref{Fig6}(c1-c3), we obtain the state-independent Heisenberg uncertainty relation in Fig. \ref{Fig6}(d), which is in good agreement with the prediction in Fig. \ref{Fig6}(b).


\section{Discussion}

For general cases, whose arbitrary group of three incompatible observables can be described as
\begin{eqnarray}
\bm{a}&=&(\cos\phi,\sin\varphi_2\sin\phi,\cos\varphi_2\sin\phi) \notag \\
\bm{b}&=&(0,\sin\varphi_1,\cos\varphi_1), \ \bm{c}=(1,0,0),
\label{Eq10s}
\end{eqnarray}
with $\varphi_1,\varphi_2\in[0,2\pi]$ and $\phi\in[0,\pi]$, the corresponding approximate observables cannot be simply given. Nevertheless, their triplewise joint measurements should satisfy Eq. (\ref{Eq1}), implying that the essence of the Heisenberg uncertainty relations in the general cases has been reflected by the three special cases considered in above sections.

The triplewise joint measurements demonstrated  here are concerning much more complex situation and essential disposition  than the pairwise counterparts considered previously \cite {PRA-89-012129,SA-2-e1600578}. It is quite interesting to note that for the case when only two incompatible observables are involved, there is no possibility to  observe a simultaneous change for two different uncertainties denoted by Wesserstein distances, which is a meticulous obligation to the realization of the fundamental limitation imposed by the Heisenberg uncertainty in this case, i.e., when one observable is more precisely measured, the other is more perturbed and fuzzy. In contrast, the Heisenberg uncertainty by triplewise joint measurements works under the condition of no simultaneous change for all the uncertainties regarding the three incompatible observables. As a result, two uncertainties, e.g., $\Delta(\mathcal{B},\mathcal{E})$ and $\Delta(\mathcal{C},\mathcal{F})$ in Fig. \ref{Eq5}, could perhaps behave identically. The idea provided in the current work is more fascinating to furnish a new possibility in the precision measurement, such that we may have two incompatible observables precisely measured by sacrificing the precision of the third incompatible observable, which is legitimately a quintessential realization of essence of the Heisenberg uncertainty with enhanced degree of computation capability.
From the data presented in Figs. \ref{Fig4}-\ref{Fig6}, we can envision a magnificent agreement between the experimental observation and the numerical prediction. Deviations of the measurement data from the numerical curves appear in some figures regarding Wesserstein distances, which are due to accumulation of the errors regarding the actual experimental data points. For example, the error bars in Fig. \ref{Eq5}(c1-c3) and in Fig. \ref{Eq5}(d) are, respectively, four and six times as long as in Fig. \ref{Eq5}(c4-c9), reflecting the maximum possible deviation of statistics in measuring four and six variables.
Nevertheless, all the deviations in our observation are non-sequential and completely within the allowed regions of experimental errors, as denoted by the error bars.

A clear understanding of the operational imperfections is essential for the precise and accurate experimental observation of the uncertainty relations. In our experiment, the resultant typical errors maybe due to any of the following aspects. (1) Imperfection exists in initial-state preparation and the final-state detection since the occupation probability of the initial state is about 98.9(2.3)$\%$ and the detection error yields a mean deviation of 0.22(8)$\%$. (2) Thermal phonons from the radial direction creates an additional dephasing effect on qubit system, yielding the depasing time of 0.24(15) ms. (3) Statistical errors always exist due to quantum projection noise. The imprecision in (1) and (2), as experimentally determined errors, can be partially calibrated by practical methods \cite{NJP-14-053053}, whereas the statistical errors are resorted into the standard deviation indicated by error bars.
Besides, the error bars also include the influence from the fluctuation in the measurements due to instability of the laser power and the magnetic field.
The fluctuation contributes ~2$\%$ error in initial-state preparations and qubit operations.

\section{Conclusion}

With progressive measurement techniques towards ultimate quantum limit, Heisenberg uncertainty relations have again become an attractive topic since it indicates that quantum mechanics is
inherently nondeterministic that there exist experiments whose outcomes cannot be predicted with arbitrary precision. Due to the same reason, the uncertainty relations also help for further understanding the foundation of quantum mechanics, such as the deeper reason for nonlocality \cite{nonlocal1,nonlocal2} and entanglement \cite{nonlocal3}. Besides, the uncertainty relations constitute an important ingredient of many device-independent security proofs \cite{prx2013,qinf1} and quantum memory \cite{qinf2}, which belong to technological foundation in quantum information processing.

We have demonstrated by a single ultracold trapped ion system the first exploration of Heisenberg uncertainty relations from the triplewise joint measurement of three incompatible observables in a pure quantum system. As only a single qubit is involved, the three incompatible observables formed by Pauli operators constitute a complete description of the quantum system, whose uncertainty relations would  bring more insight, than the counterpart regarding two incompatible observables, into the understanding of quantum characteristics at the fundamental level of a single spin.
In practice, our methods and results for the triplewise joint measurements are of distinct and more appealing characteristics in contrast to the previous investigations in the related domain, which intriguingly enriches our knowledge and give a new perception to make a more vivid and realistic understanding of the uncertainty relations. Our work is anticipated to be definitely of peculiar importance and potential applications in understanding and further development of quantum information science.

\section*{Acknowledgments}
LLY and MF appreciate insightful discussion with Paul Busch before his pass-away. This work was supported by National Key Research $\&$ Development Program of China under grant No. 2017YFA0304503, by National Natural Science Foundation of
China under Grant Nos. 11835011, 11804375, 11734018, 11674360 and 61862014, by the Strategic Priority Research Program of the Chinese
Academy of Sciences under Grant No. XDB21010100, by Guangxi Key Laboratory of Trusted Software under Grant No. kx201505£¬and  the Program for Innovative Research Team of Guilin University of Electronic Technology. K.R. acknowledges thankfully support from CAS-TWAS president's fellowship. KR and TPX contributed equally to this work.

\appendix

\section{Solution to Eq. (\ref{Eq9})}

In order to numerically solve Eq. (\ref{Eq9}), we transform the constrained extreme-value problem into the unconstrained one, that is, the optimal problem in Eq. (\ref{Eq9}) is changed to the following,
\begin{eqnarray}
\tilde{\Delta}_{lb}&=&\min_{(\bm{d},\bm{e},\bm{f})\in \Xi}  \{  2(|\bm{a}-\bm{d}|+|\bm{b}-\bm{e}|+|\bm{c}-\bm{f}|)  \notag  \\
&&+N_p\max(0,\sum_{k=0}^3|\bm{\Lambda}_k-\bm{\Lambda}_{\text{FT}}|-4) \},
\label{Eq12}
\end{eqnarray}
where $N_p$ denotes the large penalty factor for the case $\sum_{k=0}^3|\bm{\Lambda}_k-\bm{\Lambda}_{\text{FT}}|> 4$. However, our purpose is just to obtain the lower bound $\Delta_{lb}$, which usually appears at the boundary $\sum_{k=0}^3|\bm{\Lambda}_k-\bm{\Lambda}_{\text{FT}}|=4$. Thus we can consider the following form,
\begin{eqnarray}
\tilde{\Delta}_{lb}&=&\min_{(\bm{d},\bm{e},\bm{f})\in \Xi} \{  2(|\bm{a}-\bm{d}|+|\bm{b}-\bm{e}|+|\bm{c}-\bm{f}|)  \notag  \\
&&+N_p(\sum_{k=0}^3|\bm{\Lambda}_k-\bm{\Lambda}_{\text{FT}}|-4)^2 \}.
\label{Eq13}
\end{eqnarray}

\begin{figure}[hbtp]
\centering {\includegraphics[width=8 cm, height=6.0 cm]{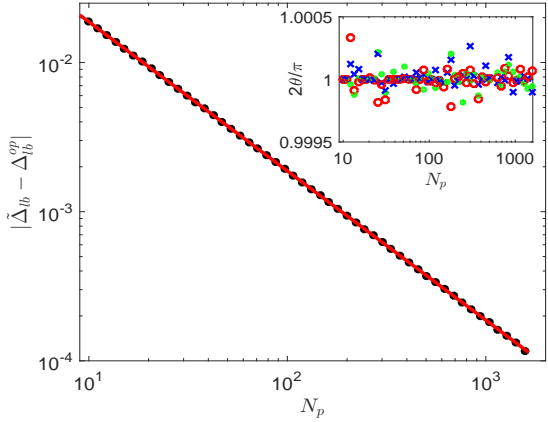}}
\caption{(Color online) Relation between $N_p$ and $|\tilde{\Delta}_{lb}-\Delta_{lb}^{op}|$ for three orthogonal incompatible observables, where we choose $ \bm{a}=(0,0,1),\bm{b}=(0,1,0),\bm{c}=(1,0,0) $ and the initial iteration condition for their approximation are $\bm{d}=\bm{a}/2,\bm{e}=\bm{b}/2,\bm{f}=\bm{c}/2$. Numerical calculation shows that for all the $N_p$ the optimal approximations converge to $ \bm{d}=(1/\sqrt{3},0,0),\bm{e}=(0,1/\sqrt{3},0),\bm{f}=(0,0,1/\sqrt{3})$ and the optimal value $\tilde{\Delta}_{lb}$ converges to $\Delta^{op}_{lb}= 2\sqrt{3}(\sqrt{3}-1)$. Dots are obtained by numerical calculation and the line denotes an inverse function fitting, i.e., $|\tilde{\Delta}_{lb}-\Delta_{lb}^{op}|=0.1876/N_p$. Inset: the three angles between two of the optimal approximations $\bm{d}$, $\bm{e}$ and $\bm{f}$, implying that they are mutually orthogonal.}
\label{Fig2}
\end{figure}

The penalty factor in Eqs. (\ref{Eq12}) and (\ref{Eq13}) is relevant to the accuracy of the numerical calculation for $\Delta_{lb}$, that is, a larger value of $N_p$ makes $\tilde{\Delta}_{lb}$ closer to $\Delta_{lb}$, i.e., $ \lim_{N_p\rightarrow\infty}\tilde{\Delta}_{lb}=\Delta_{lb}$. However, the larger value of $N_p$ makes the solution more time-consuming. As a result, we have to find a modest value of $N_p$ satisfying $|\Delta_{lb}-\tilde{\Delta}_{lb}|\leq\varepsilon$ with $\varepsilon$ being a small positive number. Fig. \ref{Fig2} demonstrates the numerical assessment for the relation between $N_p$ and $|\Delta_{lb}-\tilde{\Delta}_{lb}|$, which asymptotically obeys $|\Delta_{lb}-\tilde{\Delta}_{lb}|\sim N_p^{-1}$.

Moreover, an inlaid optimization process is involved in Eq. (\ref{Eq13}) for solving the FT vector $\bm{\Lambda}_{\text{FT}}$ of the four different vectors. This inlaid optimization process brings in a numerical obstacle for fast solving Eq. (\ref{Eq13}), which makes the solution time-consuming and inaccurate. Therefore, we have to only consider some special cases in our experimental demonstration of Heisenberg uncertainty relations for the three incompatible observables.

\section{Details for three orthogonal incompatible observables}

\begin{figure}[hbtp]
\centering {\includegraphics[width=8.5 cm, height=4.0 cm]{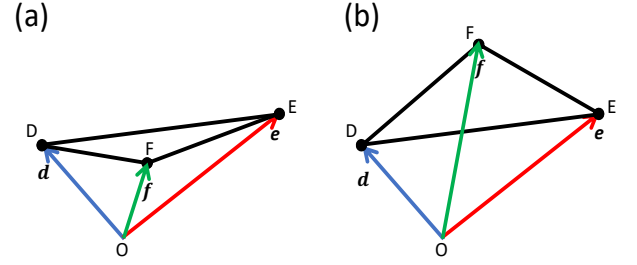}}
\caption{(Color online) Schematic diagram for different configurations constituted by the coplanar vectors $\bm{d}$, $\bm{e}$ and $\bm{f}$.}
\label{Fig7}
\end{figure}

\begin{figure*}[hbtp!]
\centering {\includegraphics[width=8.5 cm, height=6.0 cm]{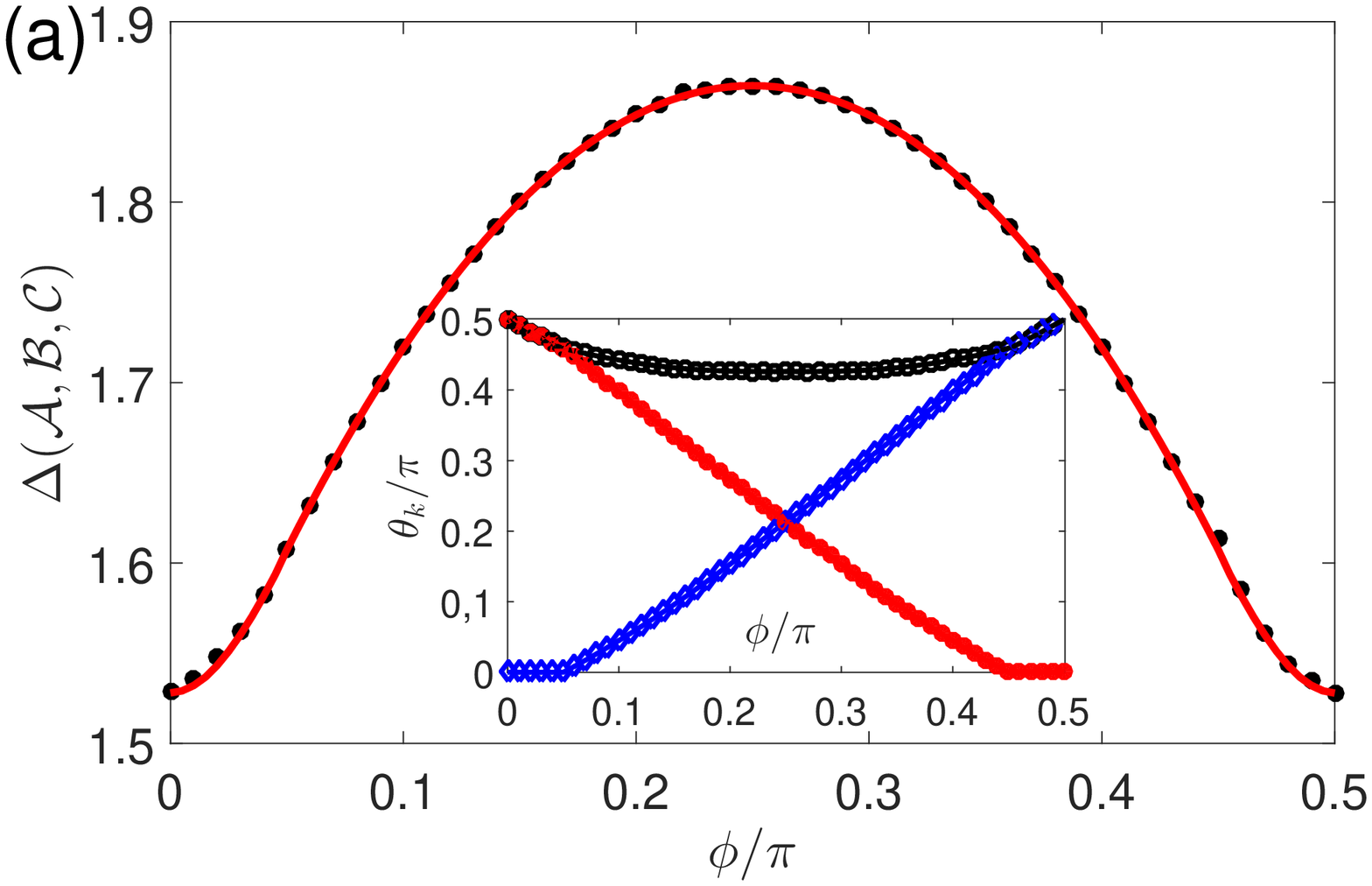}}
\centering {\includegraphics[width=8.5 cm, height=6.0 cm]{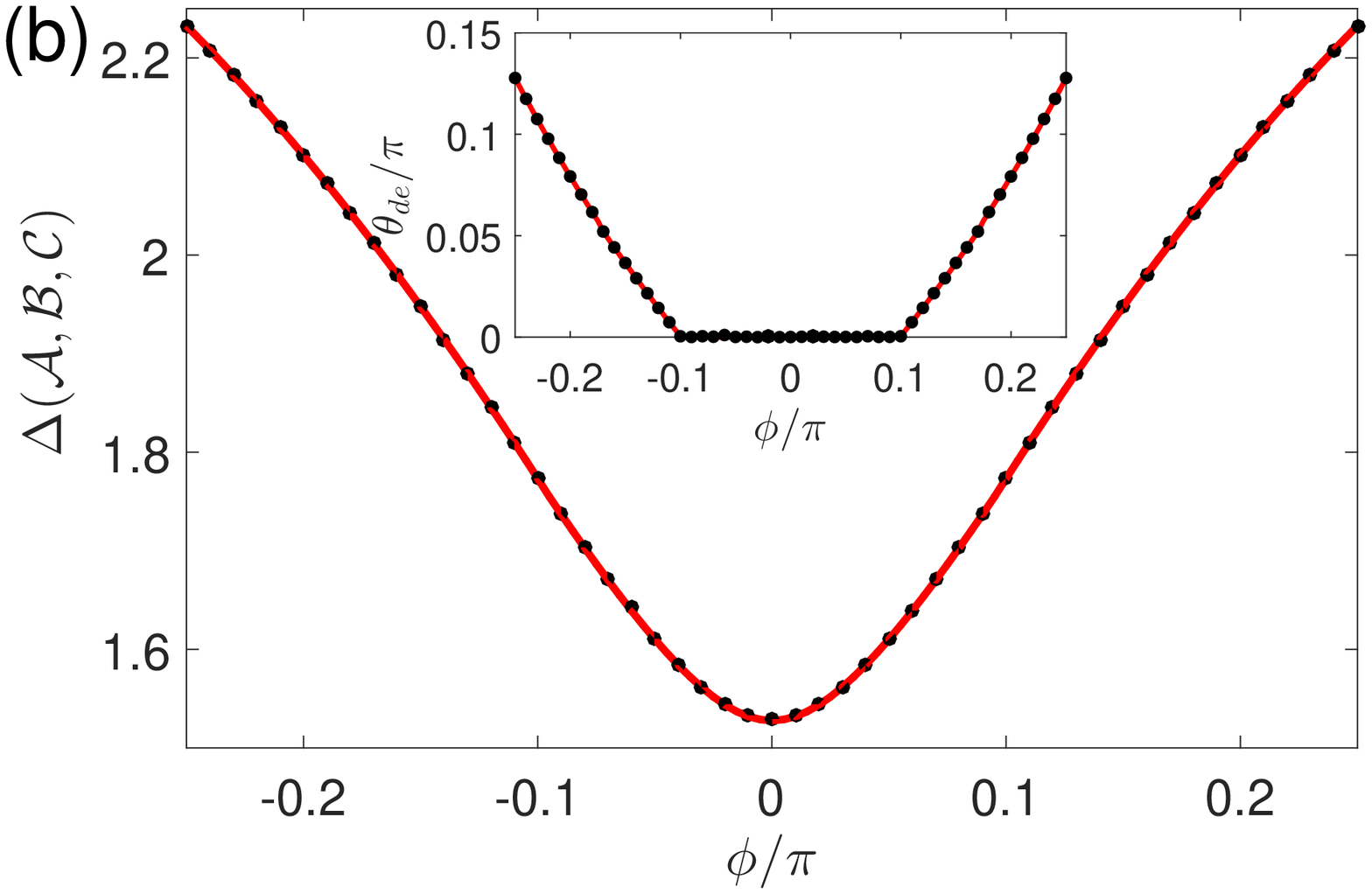}}
\caption{(Color online) Numerical results of the Heisenberg uncertainty relation obtained by the original equation, i.e., Eq. (\ref{Eq13}) (dots) and the simplified equations, i.e., Eq. (\ref{Eq21}) in (a) and Eq. (\ref{Eq24}) in (b). In (a) we choose $ \bm{a}=(0,0,1),\bm{b}=(0,1,0),\bm{c}=(0,\cos\phi,\sin\phi) $ and the initial iteration condition for their approximation is chosen as $ \bm{d}=\bm{a}/2,\bm{e}=\bm{b}/2,\bm{f}=\bm{c}/2$. In (b) we choose $ \bm{a}=(0,0,1),\bm{b}=(0,\sin\phi,\cos\phi),\bm{c}=(1,0,0) $ and the initial iteration condition for their approximation is chosen as $ \bm{d}=(2\bm{a}+\bm{b})/3,\bm{e}=(\bm{a}+2\bm{b})/2,\bm{f}=\bm{c}/2$. Inset of (a): the intersection angles reagrding the optimal approximations $\bm{d}$, $\bm{e}$ and $\bm{f}$ that $\theta_{1}$ denotes the intersection angle between $\bm{d}$ and $\bm{e}$ (black dots), $\theta_{2}$ (red dots) for $\bm{d}$ and $\bm{f}$ and $\theta_3$ (blue dots) for $\bm{e}$ and $\bm{f}$, satisfying $|\bm{d}\times\bm{e}\cdot\bm{f}|<10^{-5}$. Inset of (b): $\theta_{de}$ denotes the intersection angle between the optimal approximations $\bm{d}$ and $\bm{e}$. Other intersection angle $\theta_{df}$ ($\theta_{ef}$) between $\bm{d}$ and $\bm{f}$ ($\bm{e}$ and $\bm{f}$) satisfies $|\theta_{df}-\pi/2|<10^{-3}$ ($|\theta_{ef}-\pi/2|<10^{-3}$).}
\label{Fig3}
\end{figure*}

The analytical result can be found in the case of three orthogonal incompatible observables $\mathcal{A},\mathcal{B},\mathcal{C}$. In comparison with the case of two incompatible observables with the optimal approximations along their corresponding directions \cite{njp-19-063032,SA-2-e1600578}, we consider that the three incompatible observables under our consideration should behave similarly. Therefore, we employ the orthogonal triplewise-jointly-measurable observables as the approximations of the orthogonal incompatible observables $ \mathcal{A},\mathcal{B},\mathcal{C}$, which means $\bm{d},\bm{e},\bm{f}$ obeying the constraint condition in Eq. (\ref{Eq3}). Thus we assume,
\begin{equation}
\bm{d}=k\sin\varphi\sin\phi \bm{a},\ \bm{e}=k\cos\varphi\sin\phi \bm{b},\ \bm{f}=k\cos\phi\bm{c},
\label{Eq14}
\end{equation}
with $k\leq 1$, $ \varphi\in [0,2\pi] $ and $\phi\in [0,\pi] $. In this case, the lower bound is rewritten as
\begin{eqnarray}
\Delta_{lb1}^{op}&=&\min_{k,\varphi,\phi} 2[3-k(\sin\varphi\sin\phi+\cos\varphi\sin\phi+\cos\phi)] \notag \\
&=& 2\sqrt{3}(\sqrt{3}-1),
\label{Eq15}
\end{eqnarray}
where the minimum is obtained at $k=1$, $\varphi=\pi/4$ and $\phi=\arccos\sqrt{1/3}$.  To justify $\Delta_{lb1}^{op}$ obtained by Eq. (\ref{Eq15}) to be the real lower bound for three orthogonal incompatible observables $\mathcal{A},\mathcal{B},\mathcal{C}$, i.e., $\Delta_{lb1}^{op}=\Delta_{lb1}$, we provide in Fig. \ref{Fig2} a proof by numerical calculation, which shows a rapid decrease of $|\tilde{\Delta}_{lb1}-\Delta_{lb1}^{op}|$ with the increase of $N_p$. So we have $\lim_{N_p\rightarrow\infty}\tilde{\Delta}_{lb1}=\Delta^{op}_{lb1}$. On the other hand, we also have $ \lim_{N_p\rightarrow\infty}\tilde{\Delta}_{lb1}=\Delta_{lb1}$, which indicates that we can take $\Delta_{lb1}^{op}$ as $\Delta_{lb1}$. Therefore, for three orthogonal incompatible observables, the uncertainty relation can be stated as
\begin{equation}
\Delta(\mathcal{A},\mathcal{B},\mathcal{C})\geq 2\sqrt{3}(\sqrt{3}-1).
\label{Eq16}
\end{equation}
For experimental demonstration of Eq. (\ref{Eq16}), we choose $\bm{d}$, $\bm{e}$ and $\bm{f}$ to satisfy Eq. (\ref{Eq14}). Under the condition that the FT vector always satisfies $\bm{\Lambda}_{\text{FT}}=0$, the objective function in Eq. (\ref{Eq13}) is reduced to
\begin{eqnarray}
\bar{\Delta}_{lb1}&=&\min_{(\bm{d},\bm{e},\bm{f})\in \Xi} \{  2(|\bm{a}-\bm{d}|+|\bm{b}-\bm{e}|+|\bm{c}-\bm{f}|) \notag  \\
&&+\bar{N}_p[ g_1(\bm{d},\bm{e},\bm{f})+g_2(\bm{d},\bm{e},\bm{f})] \}, \quad
\label{Eq17}
\end{eqnarray}
where the triplewise jointly measurable and orthogonal constraints are $g_1(\bm{d},\bm{e},\bm{f})=(|\bm{d}|^2+|\bm{e}|^2+|\bm{f}|^2-1)^2$ and $g_2(\bm{d},\bm{e},\bm{f})=(\bm{d}\cdot\bm{e})^2+(\bm{d}\cdot\bm{f})^2+(\bm{e}\cdot\bm{f})^2$, respectively. Obviously, $\lim_{\bar{N}_p\rightarrow\infty}\bar{\Delta}_{lb1}=\Delta_{lb1}^{op}$ is satisfied. By changing the parameters $\phi$ and $\varphi$, we can experimentally reach the lower bound as stated in Eq. (\ref{Eq16}).

\section{Details for three coplanar incompatible observables}

Following the idea in above section, we assume that, for three coplanar incompatible observables, the corresponding optimal approximations $\bm{d}$, $\bm{e}$ and $\bm{f}$ should also lie in a plane. In this case, the triplewise jointly measurable condition in Eq. (\ref{Eq1}) can be simplified. If $\bm{f}$ lies in the triangle formed by $\bm{d}$ and $\bm{e}$, the triplewise jointly measurable condition is reduced to
\begin{equation}
|\bm{d}+\bm{e}|+|\bm{d}-\bm{e}|\leq 2.
\label{Eq19}
\end{equation}
Otherwise, the triplewise jointly measurable condition is written as
\begin{equation}
|\bm{d}+\bm{e}|+|\bm{d}-\bm{f}|+|\bm{e}-\bm{f}|\leq 2.
\label{Eq20}
\end{equation}
To be concrete, we consider the original point $O$ and the vectors $\bm{d}$, $\bm{e}$ forming a triangle $\Delta_{ODE}$, as sketched in Fig. \ref{Fig7}.
If $F$ is inside $\Delta_{ODE}$, we have the area relation $S_{ODE}=S_{ODF}+S_{DEF}+S_{EOF}$. Otherwise, $F$ is outside $\Delta_{ODE}$. The objective function in Eq. (\ref{Eq13}) is thereby reduced to
\begin{eqnarray}
\bar{\Delta}_{lb2}&=&\min_{(\bm{d},\bm{e},\bm{f})\in \Xi} \{  2(|\bm{a}-\bm{d}|+|\bm{b}-\bm{e}|+|\bm{c}-\bm{f}|)  \notag  \\
&&+\bar{N}_p[g_1(\bm{d},\bm{e},\bm{f})+g_2(\bm{d},\bm{e},\bm{f})] \},
\label{Eq21}
\end{eqnarray}
where the triplewise jointly measurable constraint is $g_1(\bm{d},\bm{e},\bm{f})=(|\bm{d}+\bm{e}|+|\bm{d}-\bm{e}|-2)^2$ if $\bm{f}$ is inside the triangle $\Delta_{ODE}$, and otherwise $g_1(\bm{d},\bm{e},\bm{f})=(|\bm{d}+\bm{e}|+|\bm{d}-\bm{f}|+|\bm{e}-\bm{f}|-2)^2$. The function $g_2(\bm{d},\bm{e},\bm{f})=(\bm{d}\times\bm{e}\cdot\bm{f})^2$ works as the coplanar constraint. Compared with Eq. (\ref{Eq13}), Eq. (\ref{Eq21}) largely simplifies the numerical process and improves the accuracy of numerical calculation.

Similar to the results in above section, the lower bound $\Delta_{lb2}$ can be obtained by an appropriate value of $\bar{N}_p$. To justify Eq. (\ref{Eq21}), we have carried out numerical simulation for $|\bar{\Delta}_{lb2}-\tilde{\Delta}_{lb2}|$, as plotted in Fig. \ref{Fig3}(a). The difference between $\bar{\Delta}_{lb2}$ and $\tilde{\Delta}_{lb2}$ is smaller than $10^{-3}$, implying  $\bar{\Delta}_{lb2}=\tilde{\Delta}_{lb2}$. Moreover, we have also checked the vector product $|\bm{d}\times\bm{e}\cdot\bm{f}|< 10^{-5}$, implying $\bm{d}$, $\bm{e}$ and $\bm{f}$ being coplanar. Therefore, Eq. (\ref{Eq21}) could be reasonably employed to calculate the lower bound of the Heisenberg uncertainty relation in comparison with the experimental measurements.

\section{Details for three incompatible observables with one orthogonal to the other two}

In this case, the corresponding optimal approximations $\bm{d}$, $\bm{e}$ and $\bm{f}$ should satisfy both the orthogonal condition that one of the three observables is orthogonal to the others and the triplewise jointly measurable condition. In the case of $\bm{f}$ orthogonal to both $\bm{d}$ and $\bm{e}$, the jointly measurable condition is reduced to
\begin{equation}
|\bm{d}+\bm{e}|+|\bm{d}-\bm{e}|\leq 2\sqrt{1-|\bm{f}|^2},
\label{Eq23}
\end{equation}
whose objective function could be given by reducing Eq. (\ref{Eq13}) to
\begin{eqnarray}
\bar{\Delta}_{lb3}&=&\min_{(\bm{d},\bm{e},\bm{f})\in \Xi} \{  2(|\bm{a}-\bm{d}|+|\bm{b}-\bm{e}|+|\bm{c}-\bm{f}|)+  \notag  \\
&&\bar{N}_p[g_1(\bm{d},\bm{e},\bm{f})+g_2(\bm{d},\bm{e},\bm{f})] \},
\label{Eq24}
\end{eqnarray}
where $g_1(\bm{d},\bm{e},\bm{f})=(|\bm{d}+\bm{e}|+|\bm{d}-\bm{e}|- 2\sqrt{1-|\bm{f}|^2})^2$ and $g_2(\bm{d},\bm{e},\bm{f})=(\bm{d}\cdot\bm{f})^2+(\bm{e}\cdot\bm{f})^2$.

Similar to the above sections, the lower bound $\Delta_{lb3}$ could be obtained by choosing a reasonable large value of $\bar{N}_p$.  We have numerically checked
$|\bar{\Delta}_{lb3}-\tilde{\Delta}_{lb3}|$ as shown in Fig. \ref{Fig3}(b). The optimal approximate observables $\bm{d}$, $\bm{e}$ and $\bm{f}$ also satisfy the orthogonal relation $\bm{f}\perp\bm{d},\bm{f}$, whereas they might have intersection angles with $\bm{a}$, $\bm{b}$ and $\bm{c}$, respectively. Moreover, the FT vector of $\bm{d}$, $\bm{e}$ and $\bm{f}$ in this case can be analytically solved as
\begin{equation}
\bm{\Lambda}_{\text{FT}}=\frac{|\bm{d}+\bm{e}|-|\bm{d}-\bm{e}|}{|\bm{d}+\bm{e}|+|\bm{d}-\bm{e}|}\bm{f}.
\label{Eq33}
\end{equation}
Thus, the measurement operators of the joint measurement observables in Eq. (\ref{Eq2}) can be analytically obtained by inserting Eq. (\ref{Eq33}) into Eq. (\ref{Eq2}).

\section{Some details of experimental manipulation}

In our experiment with the single ultracold ion, the state manipulation and experimental evolution are performed by an ultra-stable narrow linewidth Ti:Sapphire 729-nm laser corresponding to linewidth (FWHM) of 7 Hz, as measured via the heterodyne beat note method with respect to another laser system. The 729-nm laser is locked to a high-finesse ultra-low expansion cavity with the long-term drift to be 0.06 Hz/s.

For producing well-ordered laser pulses in the implementation, all the lasers are controlled by the acousto-optic modulators (AOM) by passing through the AOMs before irradiating the ion. The operational systematic rf signals applied to all the AOMs are being supplied by the direct digital synthesizer (DDS) which is controlled by a field programmable gate array. The DDS functions as the phase and frequency control of all the lasers during the consecutive experimental progressions. A typical experimental sequence involves more than 300 optical pulses within a time slot of about 40 ms.

Implementing the required evolutions comprises of two kinds of physical operations. For preparing a required state, we execute carrier transitions for different time spans by the 729-nm laser pulses with the mutual phase difference of 0, $\pi$ or any other desired value. For measuring the observables, we first generate a superposition of the states $\mid\downarrow\rangle$ and $\mid\uparrow\rangle$, and then  the detection is made by applying the cooling lasers again and counting the emitted photons for 6 ms by the photon multiplier tube. Each data point of the observables in the scheme is typically measured for 20,000 times.

\end{document}